\documentclass[twocolumn,tighten]{aastex63}
\usepackage{color}
\definecolor{mygreen}{rgb}{0,0.6,0}
\definecolor{mygray}{rgb}{0.5,0.5,0.5}
\definecolor{mydarkgray}{rgb}{0.3,0.3,0.3}
\definecolor{myred}{rgb}{0.8,0,0}
\definecolor{myblue}{rgb}{0.0,0,0.9}

\usepackage{amsmath}	
\usepackage{bm}	
\usepackage{enumitem}
\usepackage{listings}

\hypersetup{
  bookmarks=false,         
  bookmarksopen=false,     
  pdffitwindow=true,       
  pdftitle={Hypercubes of AGN Tori (HYPERCAT) -- II. Resolving the Torus with Extremely Large Telescopes},    
  pdfauthor={R. Nikutta et al.},   
  colorlinks=true,         
  linkcolor=blue,          
  citecolor=blue,          
  urlcolor=blue,           
  linktocpage=true
}

\def\about         {\hbox{$\sim$}}

\newcommand\dif    {\hbox{${d}$}}

\newcommand\Lb   {\ensuremath{L_{\rm bol}}}
\newcommand\mic  {\hbox{$\mu{\rm m}$}}

\newcommand\C {\textsc{Clumpy}}        
\newcommand\D {\textsc{Dusty}}
\newcommand\HC {\textsc{Hypercat}}        

\newcommand\N    {\hbox{${\cal N}$}}  

\newcommand\No   {\hbox{${\cal N}_0$}}    
\newcommand\iv  {\hbox{$i$}} 
\newcommand\sig  {\hbox{$\sigma$}}
\newcommand\q    {\hbox{$q$}}
\newcommand\tv   {\hbox{$\tau_{\rm V}$}}  
\newcommand\Y    {\hbox{$Y$}}

\newcommand\pa   {\ensuremath{{\rm PA}}}
\newcommand\Rd   {\ensuremath{R_{d}}}

\newcommand\Sten {\ensuremath{S_{\!10}}}

\newcommand\ELT {ELT}
\newcommand\GMT {GMT}
\newcommand\JWST {JWST}
\newcommand\Keck {Keck}
\newcommand\TMT {TMT}
\newcommand\VLTI {VLTI}

\defcitealias{Nenkova+2008a}{N08a}

\defcitealias{Nenkova+2008b}{N08b}

\defcitealias{gravity2020}{G20}
\def\gravity {\citetalias{gravity2020}}

\defcitealias{Lopez-Rodriguez+2018}{LR18}
\def\loro {\citetalias{Lopez-Rodriguez+2018}}

\defcitealias{hc-paper1}{Paper I}
\def\paperone {\citetalias{hc-paper1}}

\defcitealias{hc-paper2}{Paper II}

\shorttitle{Hypercat -- II. Resolving the Torus with Extremely Large Telescopes}
\shortauthors{Nikutta et al.}

\begin{document}
\title[HYPERCAT -- II. Resolving the Torus with Extremely Large Telescopes]{Hypercubes of AGN Tori (HYPERCAT) -- II. Resolving the Torus with Extremely Large Telescopes}



\correspondingauthor{Robert Nikutta}
\email{robert.nikutta@noirlab.edu}

\author[0000-0002-7052-6900]{Robert Nikutta} 
\affiliation{NSF's NOIRLab, 950 N. Cherry Ave., Tucson, AZ 85719, USA}

\author[0000-0001-5357-6538]{Enrique Lopez-Rodriguez}
\affiliation{Kavli Institute for Particle Astrophysics and Cosmology (KIPAC), Stanford University, Stanford, CA 94305, USA}

\author[0000-0002-4377-903X]{Kohei Ichikawa}
\affiliation{Frontier Research Institute for Interdisciplinary Sciences, Tohoku University, Sendai 980-8578, Japan}
\affiliation{Astronomical Institute, Tohoku University, Aramaki, Aoba-ku, Sendai, Miyagi 980-8578, Japan}

\author[0000-0003-4209-639X]{N. A. Levenson}
\affiliation{Space Telescope Science Institute, Baltimore, MD 21218, USA}

\author[0000-0001-7827-5758]{Christopher Packham}
\affiliation{Department of Physics \& Astronomy, University of Texas at San Antonio, One UTSA Circle, San Antonio, TX 78249, USA}
\affiliation{National Astronomical Observatory of Japan, 2-21-1 Osawa, Mitaka, Tokyo 181-8588, Japan}

\author[0000-0002-6353-1111]{Sebastian F. H\"{o}nig}
\affiliation{School of Physics \& Astronomy, University of Southampton, Southampton SO17 1BJ, United Kingdom}

\author[0000-0001-6794-2519]{Almudena Alonso-Herrero}
\affiliation{Centro de Astrobiolog\'{\i}a (CAB, CSIC-INTA), ESAC Campus, Villanueva de la Ca\~nada, E-28692 Madrid, Spain}

\begin{abstract}
\noindent
  Recent infrared interferometric observations revealed sub-parsec
  scale dust distributions around active galactic nuclei (AGNs). Using
  images of \C\ torus models and NGC~1068 as an
  example, we demonstrate that the near- and mid-infrared nuclear
  emission of some nearby AGNs will be resolvable in direct imaging
  with the next generation of 30~m telescopes, potentially breaking
  degeneracies from previous studies that used integrated spectral
  energy distributions of unresolved AGN tori.
  To that effect we model wavelength-dependent point spread functions
  from the pupil images of various telescopes: James Webb Space
  Telescope, Keck, Giant Magellan Telescope, Thirty Meter Telescope,
  and Extremely Large Telescope. We take into account detector pixel
  scales and noise, and apply deconvolution techniques for image
  recovery.
  We also model 2D maps of the 10-\mic\ silicate feature strength,
  \Sten, of NGC~1068 and compare with observations. When the torus is
  resolved, we find \Sten\ variations across the image. However, to
  reproduce the \Sten\ measurements of an unresolved torus a dusty
  screen of $A_V > 9$~mag is required. We also fit the first resolved
  image of the K-band emission in NGC~1068 recently published by the
  GRAVITY collaboration, deriving likely model parameters of the
  underlying dust distribution. We find that both 1) an elongated
  structure suggestive of a highly inclined emission ring, and 2) a
  geometrically thin but optically thick flared disk where the
  emission arises from a narrow strip of hot cloud surface layers on
  the far inner side of the torus funnel, can explain the
  observations.
\end{abstract}


%
\section{Introduction}
\setcounter{footnote}{9}
\noindent

The thermal emission from the central region of an active galactic
nucleus (AGN) depends spectrally and spatially on the distribution of
dust close to the accreting black hole. A central dusty ``torus''
unifies disparate AGN classes \citep{KrolikBegelman1988,
  Antonucci1993, UrryPadovani1995}, and more specifically, a toroidal
distribution of clouds generally accounts for characteristic observed
features \citep[e.g.,][]{Nenkova+2002, Nenkova+2008a, Nenkova+2008b,
  Hoenig+2006, Schartmann+2008}.

The central AGN tori are relatively small in the near-infrared (NIR)
and mid-infrared (MIR), on scales of a few parsecs
\citep[e.g.,][]{Jaffe+2004, Nenkova+2008b, RamosAlmeida+2011,
  Tristram+2014}, so not currently resolvable in direct images. For
nearby AGNs the only spatially resolved observations of the inner few
parsec come from interferometers such as VLTI/GRAVITY and VLTI/MIDI in
NIR and MIR, and the Atacama Large Millimeter / submillimeter Array
(ALMA) in the submillimeter regime. Especially in the IR the baseline
coverage is comparatively poor, with just two to four telescopes
sampling the Fourier space of spatial frequencies. Image
reconstruction is either impossible, or relies heavily on modeling of
the sampled brightness distribution. However, 2D images, as opposed to
1D visibilities or 1D spectral energy distributions (SED), are needed
to break possible degeneracies in the IR radiative transfer
\citep{Vinkovic+2003}. The upcoming generation of extremely large
telescopes (Giant Magellan Telescope, GMT; Thirty Meter Telescope,
TMT; and Extremely Large Telescope, ELT) all have the potential of
resolving the IR emission in nearby AGNs with their single-aperture
mirrors (we use ``single-dish'' in the text for simplicity, but all
these telescopes have either segmented mirrors or comprise several
circular mirrors arranged in a fixed pattern).

Perhaps surprisingly, at mid-IR wavelengths ``polar'' emission is
frequently observed \citep[e.g.,][]{Hoenig+2012, Hoenig+2013,
  Burtscher+2013, Lopez-Gonzaga+2016b, Leftley+2018}, extending
perpendicular to the equatorial axis of the obscuring torus. Using
simulations we can make progress to determine whether the torus
produces this emission, or if additional material is required. The
picture is complicated by the fact that the near- to mid-IR emission
morphology does not necessarily follow the 3D dust distribution (see
results in \citet[][henceforth Paper I]{hc-paper1}).
Motivated by the situation above, in \paperone\ we have demonstrated
that the polar elongation could be reproduced under specific
conditions by utilizing the set of \C\ model images and dust maps
together with the accompanying \HC\ software.\footnote{The \HC\ models
  and software are available at \url{https://www.clumpy.org/images/}}

While we perform all analyses in this paper using the AGN torus models
produced by \C, we note that in principle any other image-generating
models could be used in conjunction with the \HC\ software. With the
\C\ models we hold several parameters fixed, e.g., the dust
composition, grain size distribution, sublimation temperature,
spectral shape of the illuminating source, and the functional
prescription of the ``softness'' of the polar torus edge (we use a
Gaussian form).

Other models might make different assumptions about the parameters,
and about the geometry itself. For instance,
\citet{Hoenig_Kishimoto_2010} studied several dust chemical
compositions and grain size distributions, and improved handling of
the diffuse radiation field in the torus, to investigate their
influence on the shape of spectral energy
distributions. \citet{Siebenmorgen+2015} introduced a two-phase medium
and fluffy (rather than spherical) dust grains in their AGN torus
model to explain the observed range of silicate feature strengths and
wavelengths of the peak emission. \citet{Stalevski+2016} varied
several parameters of their own two-phase Monte Carlo model to
investigate the relation between $\rm L_{torus}/L_{AGN}$ and the dust
covering factor. \citet{Hoenig_Kishimoto_2017} distributed the bulk of
dusty clouds in their revised CAT3D model along the polar axis of the
AGN system to help explain the observed polar elongation of the MIR
emission in several nearby AGNs. Each such model (and others not
mentioned here explicitly) is motivated by answering different open
questions.

We also note that to smoothly interpolate images in several orthogonal
dimensions (the free model parameters; seven in the case of \C), the
changes between model images from one set of parameter values to a
neighboring set ought to be smooth. This is potentially a challenge
for Monte Carlo based models \emph{if} the cloud configuration between
model realizations is allowed to change (i.e., if the cloud realization
is randomized each time). While \C\ eponymously simulates clumpy AGN
tori, it differs from Monte Carlo models in that the model image
resulting from its radiative transfer is by construction the mean of
an infinite number of individual model realizations. Monte Carlo
models of clumpy dust distributions that generate the cloud geometry
anew for each realization could potentially overcome the challenge by
averaging the output images of many realizations with identical
parameter values; but that may prove too expensive regarding the
required compute time. Models which employ smooth dust distributions
by design, such as in, e.g., \citet{Fritz+2006}, could be used
immediately with \HC, provided they were packaged in an appropriate
$n$-dimensional hypercube.\footnote{We document in the \HC\ User Manual
  the exact requirements and provide a tool to create such hypercubes
  from a collection of image files, and can also assist colleagues
  upon request.}

Here we use \HC\ simulations to model the nearby ($D = 14.4$~Mpc) and
bright ($L_{\rm bol} = 0.15-1.5 \times 10^{45}\, {\rm erg\, s^{-1}}$)
AGN NGC~1068.
Fitting the well-populated SED \citep[][hereafter
LR18]{Lopez-Rodriguez+2018} with models generated by \C\
\citep{Nenkova+2008a, Nenkova+2008b} provides geometrical parameters
of the dust distribution. We compute the wavelength-dependent point
spread functions (PSF) of several current and future telescopes --
James Webb Space Telescope (JWST); Keck; Giant Magellan Telescope
(GMT); Thirty Meter Telescope (TMT); Extremely Large Telescope (ELT)
-- to simulate realistic single-dish observations and determine the
resolvability of nearby AGNs. Our simulations also serve as predictions
of the image observations from future facilities with higher spatial
resolution.
The 10-\mic\ silicate feature provides a valuable diagnostic of the
dust distribution, especially when observed using an integral field
unit (IFU) for spatially and spectrally resolved image cubes, which we
also simulate.
Finally, we directly fit the recent 2.2-\mic\ image of NGC~1068
released by the \citet[][hereafter G20]{gravity2020}, deriving likely
physical and geometrical parameters governing the source.


\section{NGC~1068 Case Study}
\label{sec:ngc1068}
\noindent

\subsection{Torus Parameters}
\label{sec:source}
\noindent
The type~2 Seyfert galaxy NGC~1068 has been extensively studied in the
IR using single-dish \citep[e.g.,][]{Mason+2006, RamosAlmeida+2011,
  Alonso-Herrero+2011, Lira+2013, Ichikawa+2015} and interferometric
\citep[e.g.,][]{Jaffe+2004, Raban+2009, Burtscher+2013,
  Lopez-Gonzaga+2014} observations. ALMA submillimeter observations have
recently resolved the molecular torus in NGC~1068 and constrained its
diameter to $\sim\!\!12$ pc using the dust continuum at 432~\mic\ and
\hbox{CO J=6-5} \citep{Garcia-Burillo+2016}, and then found a size
$13 \times 4$ pc in the \hbox{HCN J=2-1} transition and
$12 \times 5$ pc for \hbox{HCO$^{+}$ J=3-2} \citep{Imanishi+2018}. The
torus is found to be inclined, and lopsided, at a position angle of
its mid-plane about 100--110\degr\ east of north. Including the
432~\mic\ dust continuum observations by ALMA as a constraint in the
1--432~\mic\ nuclear SED fits of NGC~1068, \citet[][hereafter
LR18]{Lopez-Rodriguez+2018} derived the \C\ torus parameters that best
reproduced the observations. We adopt these model parameters for our
present study; they are listed in
Table~\ref{table:NGC1068torusparameters}. The position angle of the
torus large axis in the plane of the sky, \pa\about 138\degr\ east of
north, was taken from 2~\mic\ polarimetric observations of the nucleus
\citep{Packham+1997, Simpson+2002, Lopez-Rodriguez+2015,
  Gratadour+2015}. The 2~\mic~polarization is thought to arise from
the absorption of aligned dust grains in the torus of NGC~1068 where
the measured polarization angle indicates the global orientation of
the equatorial plane of the torus. This \pa\ has been further
confirmed by means of magnetically aligned dust grains across the
resolved equatorial axis of the torus using ALMA polarimetric
860~\mic\ observations \citep{Lopez-Rodriguez+2020}.

\begin{figure*}
  \includegraphics[width=\textwidth,height=0.95\textheight,keepaspectratio]{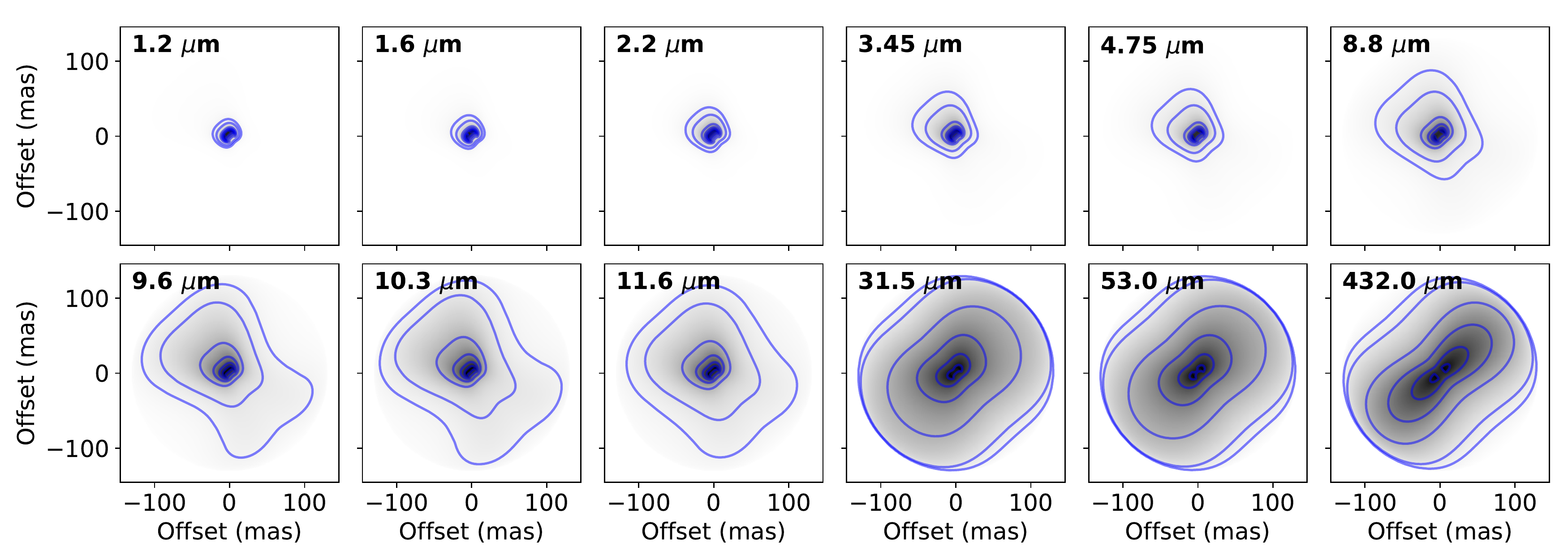}
  \caption{2D torus thermal emission images of NGC~1068 modeled with
    \C\ from 1.2~\mic\ to 432~\mic\ using the parameters from Table
    \ref{table:NGC1068torusparameters}. The morphology varies as
    function of wavelength due to changes of optical depth and dust
    temperature. Blue contours show the fractional surface brightness
    at 0.05, 0.1, 0.3, 0.5, 0.7, and 0.9 of the peak value.}
  \label{fig:figVII}
\end{figure*}

Figure~\ref{fig:figVII} shows the distribution of the thermal emission
within the 1--432~\mic\ wavelength range for a \C\ model with
parameters from Table \ref{table:NGC1068torusparameters}. The
morphology of the emission changes dramatically across the wavelength
regimes as also shown in figure 5 in \loro. For details on emission
morphology please see the companion paper \citep[][henceforth Paper
I]{hc-paper1}.
\begin{deluxetable}{llcll}
  \tablecaption{\C\ Parameters and Derived Quantities from
    the Best-fit SED Model to NGC~1068 by \loro.\label{table:NGC1068torusparameters}}
  \tablewidth{0pt} \tablehead{ \multicolumn2c{Model Parameters} &
    \phantom{000} & \multicolumn2c{Derived \& Literature Values} }
  \startdata
  \sig\ (deg) & $43^{+12}_{-15}$       &  & $R_{\rm in}$ (pc)            & $0.28^{+0.01}_{-0.01}$ \\[3pt]
  \Y           & $18^{+1}_{-1}$         &  & $R_{\rm out}$ (pc)           & $5.1^{+0.4}_{-0.4}$    \\[3pt]
  \No          & $4^{+2}_{-1}$          &  & $H$ (pc)                     & $3.5^{+1.0}_{-1.3}$    \\[3pt]
  \q           & $0.08^{+0.19}_{-0.06}$ &  & $\Lb/10^{44}$ (erg s$^{-1}$) & $5.02^{+0.15}_{-0.19}$ \\[3pt]
  \tv          & $70^{+6}_{-14}$        &  & $D^{a}$ (Mpc)                & $14.4$                 \\[3pt]
  \iv\ (deg)  & $75^{+8}_{-4}$         &  & \pa\ (deg)                  & $138$                  \\
  \enddata \tablecomments{$^{a}$At $D = 14.4$~Mpc, 1\arcsec~= 70
    pc, adopting H$_{0}$ = 73 km s$^{-1}$ Mpc$^{-1}$
    \citep{Bland-Hawthorn+1997}.}
\end{deluxetable}

\subsection{Single-dish Synthetic Observations of NGC~1068}
\label{sec:synObsNGC1068}

\subsubsection{Telescopes, Pupil Images and PSFs}
\label{sec:PSF}
\noindent
To produce synthetic observations with a given telescope and
instrument combination, the telescope PSF and the instrumental
configuration are needed. A telescope PSF can be very complex
\citep[e.g.,][]{Krist+2011, Perrin+2012} and depends on many factors;
for instance telescope aperture, telescope configuration, seeing
conditions, adaptive optics configuration, and wavelength. Our aim is
to provide the best-case scenario of a synthetic observation for a
given telescope and wavelength. We assume perfect adaptive optics, and
atmospheric conditions without sky or PSF variations. These conditions
can be refined by applying specific aberrations relevant to the
science cases and telescope/instrument configuration, but this is
outside the scope of the present work. Future releases of \HC\ may
incorporate tools to account for these conditions.

Table~\ref{table:listinstruments} summarizes the telescopes and
instruments available in \HC, which include the next generation of
30~m telescopes (\GMT, \TMT, \ELT) as well as \JWST\ and \Keck\ for
comparison. For telescopes without an instrument in a given wavelength
range the pixel scale is taken to be equal to the Nyquist sampling
($0.5 \times\lambda/D$). The FWHM is given by $\lambda/D$, except for
\JWST, whose PSFs have been modeled in detail by the instrument teams.

\begin{table*}
\caption{List of Single Dish Telescopes and Instruments Available with \HC.\label{table:listinstruments}}
\begin{tabular*}{\textwidth}{l @{\extracolsep{\fill}} clrrr}
  \hline
  \hline
  Telescope & Diameter & Instrument  & Wavelength  & Pixel scale & FWHM \\
            & (m)      &             & (\mic)      & (mas)       & (mas)\\
  \hline
  \JWST     & 6.5      & NIRCAM      &  0.6  -   5 & 31 -  63    &  31 - 162$^b$\\
            &          & NIRISS      &  0.6  -   5 & 31 -  63    &  42 - 152$^c$\\
            &          & MIRI        &    5  -  28 &      110    & 182 - 812$^d$\\
  \Keck     & 10.0     & NIRC2       &  0.9  - 5.3 & 10 -  40    & 163 - 241$^e$\\
            &          & Nyquist$^a$ &    8  -  20 & 83 - 206    & 165 - 412\phantom{$^f$}\\
  \GMT      & 24.5     & GMTIFS      &  0.9  - 2.5 &        5    &   8 -  21\phantom{$^f$}\\
            &          & Nyquist$^a$ &    3  -   5 & 12 -  21    &  25 -  42\phantom{$^f$}\\
            &          & Nyquist$^a$ &    8  -  20 & 34 -  84    &  67 - 168\phantom{$^f$}\\
  \TMT      & 30.0     & IRIS        & 0.85  - 2.5 &        4    &   6 -  17\phantom{$^f$}\\
            &          & MICHI       &    3  -   5 &     11.9    &  21 -  34\phantom{$^f$}\\
            &          & MICHI       &    8  -  12 &     27.5    &  55 -  82\phantom{$^f$}\\
  \ELT      & 39.1     & MICADO      &  0.8  - 2.5 &        4    &   4 -  13\phantom{$^f$}\\
            &          & METIS       &  2.9  -   5 &        4    &  15 -  26\phantom{$^f$}\\
            &          & METIS       &    5  -  14 &       13    &  26 -  74\phantom{$^f$}\\
  \hline
\end{tabular*}
$^a$For telescopes without an instrument in a given wavelength range, the pixel scale is set to the Nyquist sampling at the shortest wavelength. For FWHM values without a superscript, $\lambda/D$ was assumed. Otherwise the FWHM values are from:\\
$^b$\hbox{\url{https://jwst-docs.stsci.edu/near-infrared-camera/nircam-predicted-performance/nircam-point-spread-functions}}\\
$^c$\hbox{\url{https://jwst-docs.stsci.edu/near-infrared-imager-and-slitless-spectrograph/niriss-predicted-performance/niriss-point-spread-functions}}\\
$^d$\hbox{\url{https://jwst-docs.stsci.edu/mid-infrared-instrument/miri-predicted-performance/miri-point-spread-functions}}\\
$^e$\hbox{\url{https://www2.keck.hawaii.edu/inst/nirc2/ScamCompare.html}}
\end{table*}

\begin{figure*}[t]
\center \includegraphics[width=0.8\textwidth,keepaspectratio]{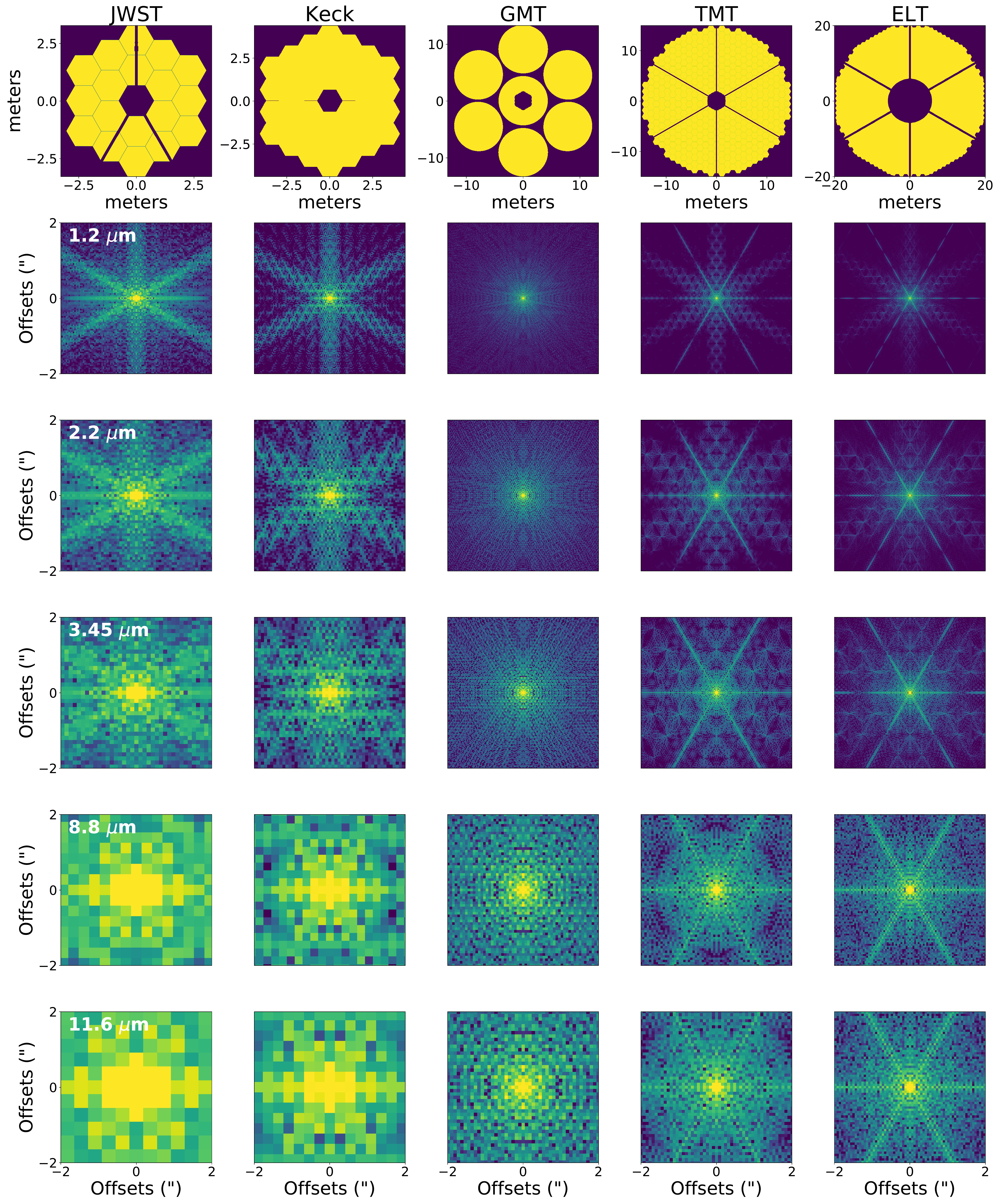}
\caption{First row shows the pupil images of the \JWST, \Keck, \GMT,
  \TMT, and \ELT\ from left to right. Their PSFs at 1.2, 2.2, 3.45,
  8.8 and 11.6 \mic~within $4 \times 4$ arcsec$^{\rm 2}$ from second
  to bottom row are shown in logarithmic scale. For all PSFs, we have
  adopted the Nyquist sampling as the pixel scale.}
\label{fig:figII}
\end{figure*}

We compute the PSF as the Fourier transform of the telescope pupil
image (see Appendix \ref{app:PSF} for details). The PSF includes the
effects of the telescope structures that interact with the science
beam. The pupil images are in units of meters, which are then
converted to arcsec using the PIXSCALE keyword in the corresponding
FITS files. For any given wavelength and telescope, the PSF has a FWHM
= $\lambda/D$. Figure \ref{fig:figII} shows the pupils and PSFs of the
telescopes from Table \ref{table:listinstruments}.

Note that \HC\ also provides two other methods to estimate the PSF:
\texttt{model-PSF}, which is a combination of a central Airy disk that
contains most of the power of the PSF, and a broad Gaussian halo that
encompasses the remaining power of the object; and \texttt{image-PSF}
where the user can provide as an image the empirical PSF of an
observed standard star for a given
telescope/instrument. \texttt{model-PSF} also accounts for Strehl
ratio to perform studies of the AO
system (Appendix \ref{app:PSF}).

To qualitatively explore the PSF for each instrument at a given
wavelength we produce monochromatic PSF images (Figure
\ref{fig:figII}) at five representative wavelengths within the
$1-13$~\mic\ wavelength range for each of the telescopes in Table
\ref{table:listinstruments}. Note that the throughput of a specific
filter may be taken into account for a more realistic PSF. Here, the
only diffractive and optical aberration component in our simulations
comes from the pupil image.

\begin{figure}
  \includegraphics[width=\columnwidth]{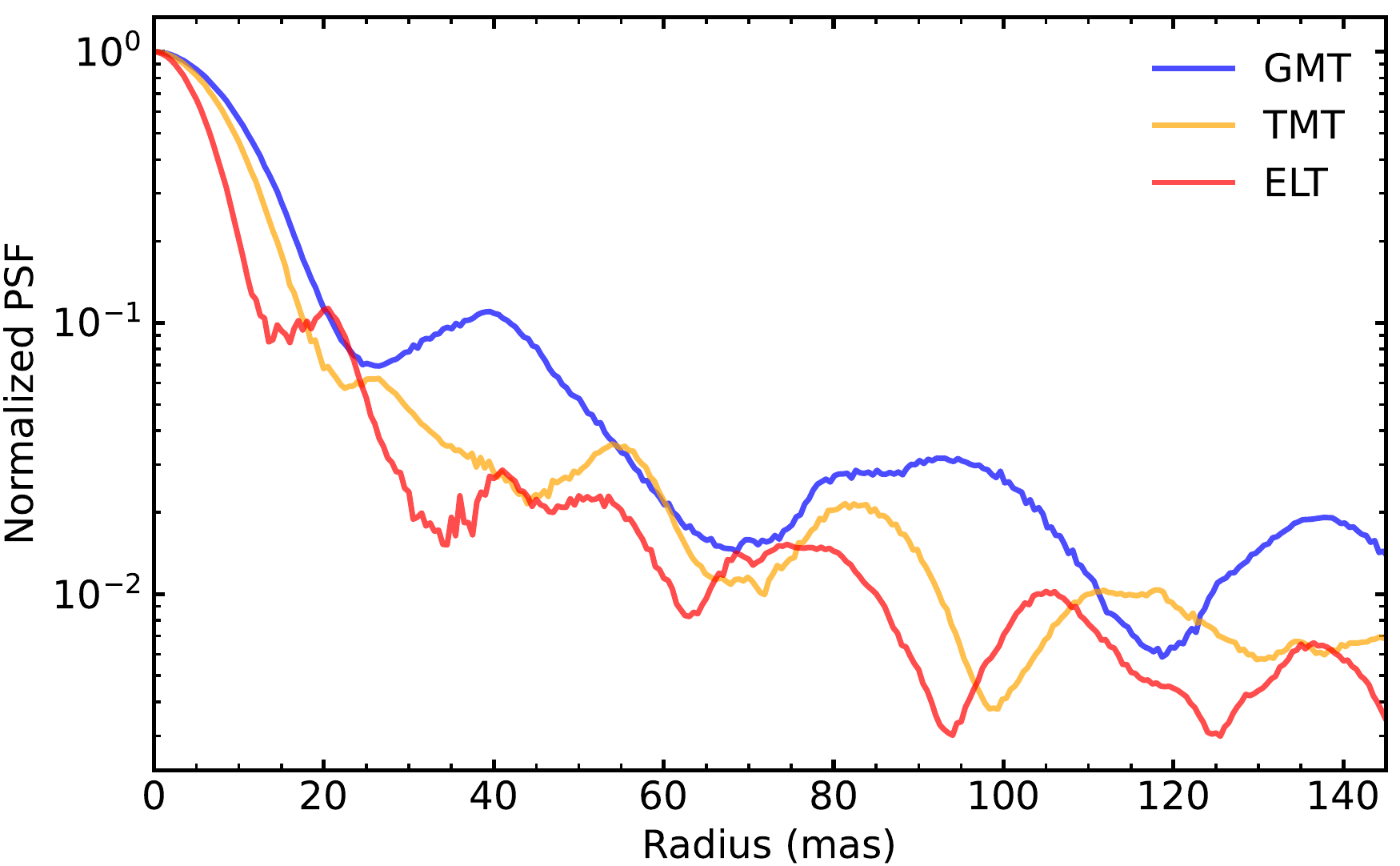}
  \caption{Radial profiles of the PSFs of \GMT, \TMT, and \ELT\ within
    the central $\sim145$ milliarcsec radius, at a wavelength of
    2~\mic.}
  \label{fig:PSFradialprofile}
\end{figure}

Figure \ref{fig:PSFradialprofile} shows the radial profiles (computed
over circular annuli) of the PSFs of \GMT, \TMT, and \ELT\ at
2.0~\mic. As expected, the `cleanest' (most Airy-like) extended power
of the PSF is provided by the \GMT, while the highest-resolution core
of the PSF is provided by the \ELT. As shown by \citet{Angel+2003},
the best extended PSF will be provided by the telescope configuration
with full disks, i.e., \GMT, over those with fitted hexagons, i.e.,
\ELT\ and \TMT. This is because the extended regions of the PSF from
fitted hexagons have more structures (i.e., `spider' shape in the
wings of the PSF) than those from full circular disks.  The
instrumental configuration will be thus of great importance when
considering specific science goals of the ground-based 30~m class
telescopes.

%

\subsubsection{Toward a Synthetic Observation}
\label{sec:SynObs}
\begin{figure*}
  \includegraphics[width=\textwidth]{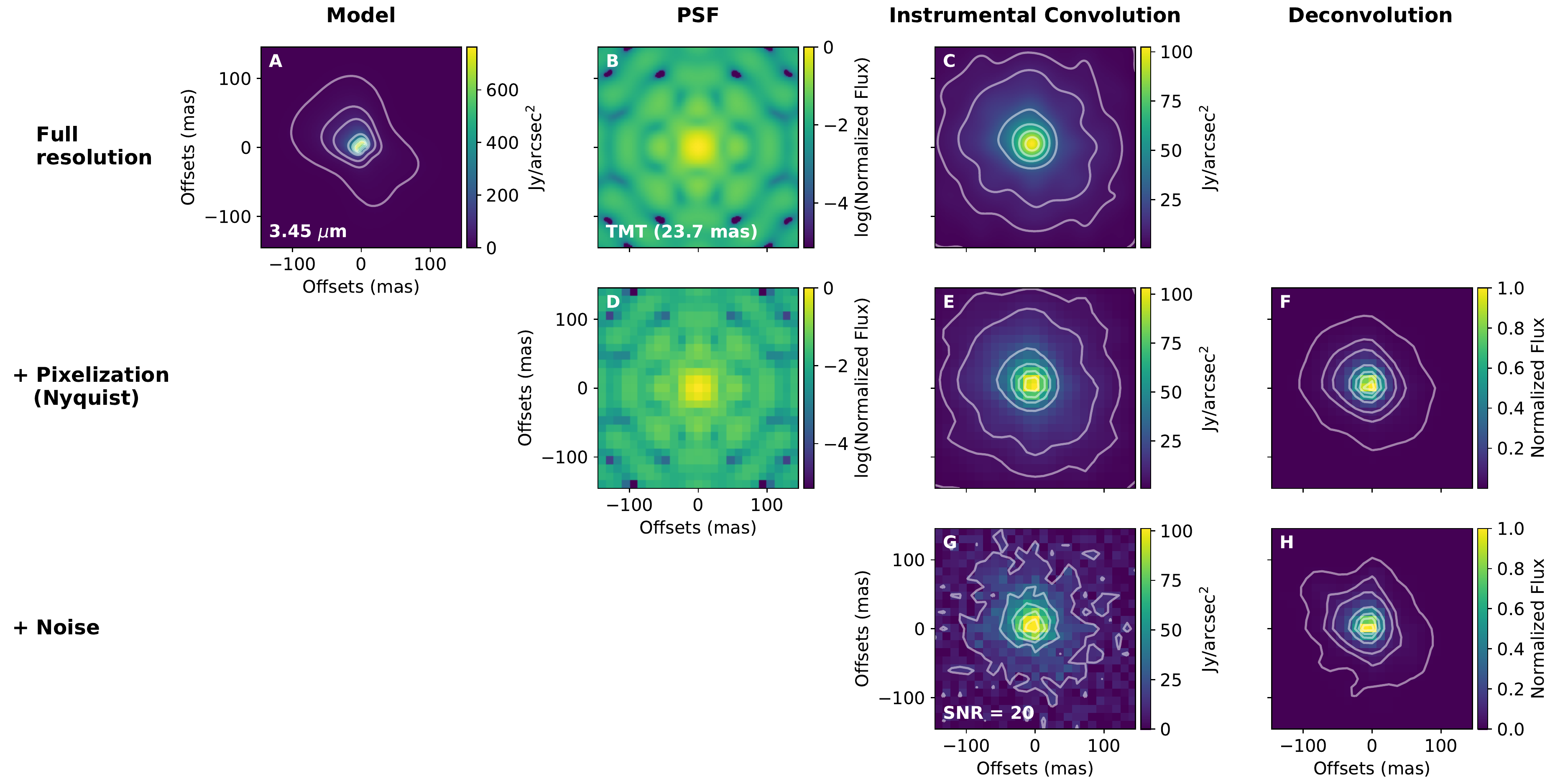}
  \caption{\HC\ step-by-step process to obtain synthetic observations
    from a 2D thermal emission map produced by using \textsc{Clumpy}
    models. From left to right the columns add operations on the input
    image by the optical system. From top to bottom the rows add
    detector degradations to the image. \emph{(A)} 3.45~\mic\ model
    image of NGC~1068 at full resolution (pixel scale 0.67 mas) using
    the parameters shown in Table \ref{table:NGC1068torusparameters},
    and applying a total flux density of 1.0~Jy obtained by \loro. All
    contours are shown at 0.01, 0.05, 0.1, 0.3, 0.5, 0.7, 0.9
    fractions of the image peak. \emph{(B)} PSF at 3.45 \mic\ computed
    using the pupil image of \TMT. The telescope angular resolution
    $\lambda/D = 23.7$~mas is printed in the lower left
    corner. \emph{(C)} Convolved image using the full resolution
    emission image (A) and PSF (B). \emph{(D)} PSF pixelated at the
    Nyquist sampling of \TMT\ at 3.45 \mic, i.e.,
    $\lambda/D/2 = 11.87$~mas. Due to discretization limits (odd
    number of pixels is enforced by \HC), the effective pixel scale is
    11.64~mas. \emph{(E)} Convolved image of NGC~1068, pixelated at
    the same effective Nyquist sampling. \emph{(F)} Deconvolved image
    using the pixelated image (E) and PSF (D). \emph{(G)} Gaussian
    noise with SNR = 20 at the peak pixel was applied to the pixelated
    image of the thermal emission, E. \emph{(H)} Deconvolved image
    using the pixelated PSF (D) and noisy image (G).}
  \label{fig:stepbystep}
\end{figure*}

\noindent
Single-dish observations are degraded by several processes
(Appendix~\ref{app:synobs}). The fundamental loss of resolution is due
to convolution of the incoming wavefront with the PSF of the
telescope.
The PSF-convolved image is then downsampled to the discrete pixel
sizes of the detectors. The PSF convolved and pixelated image (step E
in Figure \ref{fig:stepbystep}) is then flux calibrated. Specifically,
the total flux of the model is equal to the flux density, in units of
'Jy', from a known observation at the highest spatial resolution at a
given wavelength.  For our example, we used a flux density of 1 Jy at
3.45 $\mu$m by \loro.  Finally, noise due to, for instance,
fluctuations in detector readout, atmospheric emission and
transmission is taken into account (Appendix \ref{app:noise}). Note
that the final synthetic observations have a background set to zero,
background subtraction is not required.  Figure \ref{fig:stepbystep}
shows this procedure using the 2D \C\ torus image of NGC~1068 and the
PSF of \TMT\ at 3.45 \mic.
These synthetic observations can be directly compared with real data.
The rightmost column in Figure~\ref{fig:stepbystep} shows the
deconvolved images with and without noise. The improvement in
detecting extended emission by deconvolving the noisy image is very
conspicuous.

\subsubsection{Results}
\label{sec:synObsNGC1068}

\begin{figure*}[t]
  \includegraphics[width=\textwidth,height=0.95\textheight,keepaspectratio]{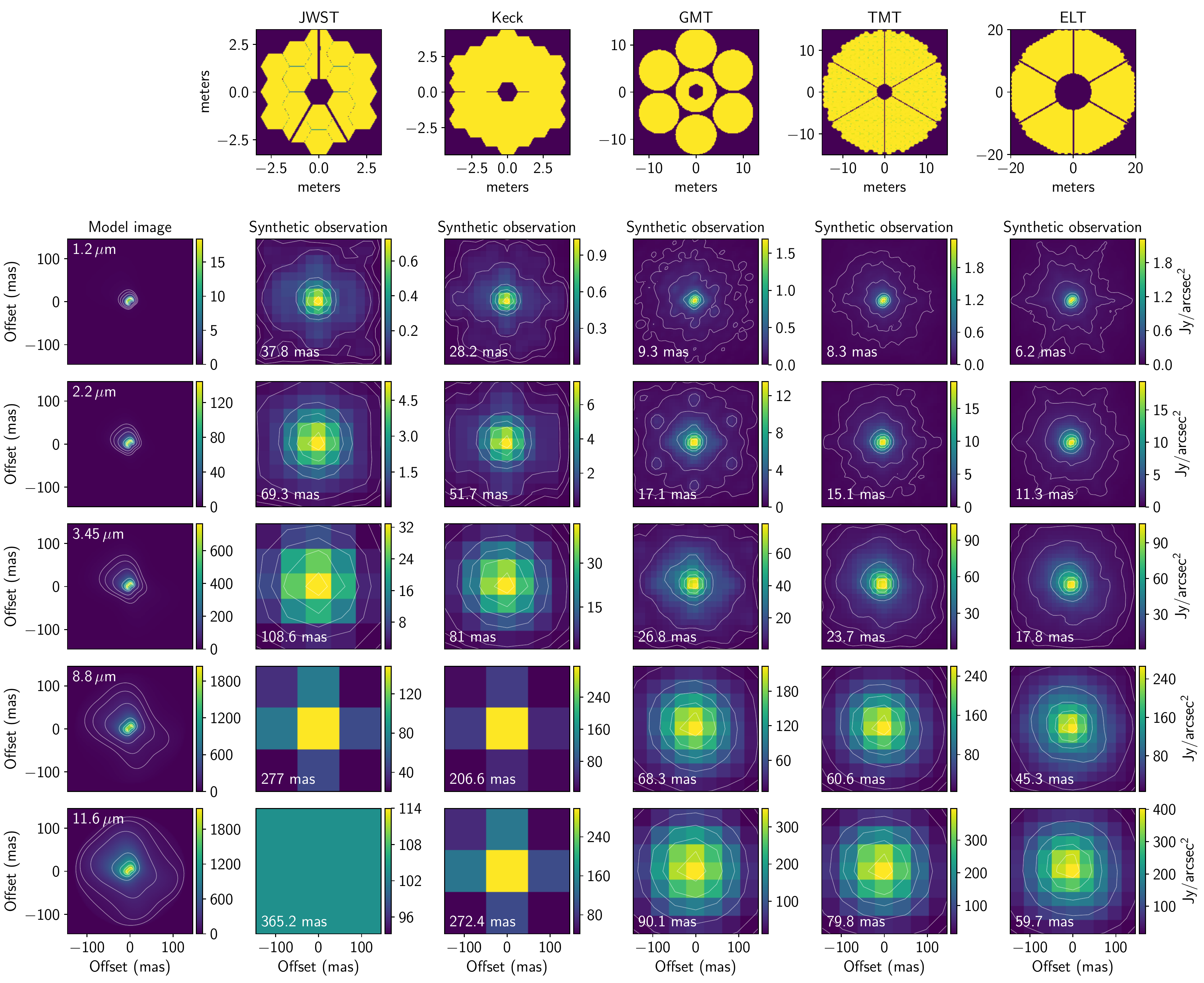}
\caption{First row shows the pupil images of the \JWST, \Keck, \GMT,
  \TMT\ and \ELT\ from left to right. From the second row, the first
  column shows the 2D \C\ torus image of NGC~1068, and columns 2-6
  show the synthetic observations for each telescope, for several
  wavelengths (rows 2-6). In all cases, model image and synthetic
  observations are flux calibrated and pixelated to Nyquist
  sampling. Contours are shown at levels of 0.03, 0.05, 0.1, 0.3, 0.5,
  0.7, 0.9 times the peak of the image, except for the \JWST\ and
  \Keck\ at 8.8 and 11.6 $\mu$m because the pixelscale is similar to
  the FOV of the image. Each panel shows the angular resolution at the
  given wavelength estimated as $\lambda/D$.}
\label{fig:figI}
\end{figure*}

Figure~\ref{fig:figI} shows the synthetic observations for each
telescope in the 1.2-11.6 \mic~wavelength range. They were obtained
following the steps shown in Figure~\ref{fig:stepbystep} through step
E, i.e., they are flux-calibrated and pixelated, but noise-free and not
deconvolved.

An outstanding result is that all the upcoming generation of 30~m
class telescopes will be able to spatially resolve the dust emission
distribution of the torus around NGC~1068 in the $1-12$ \mic\
wavelength range. Although the 6.5-m \JWST\ telescope has the most
sensitive capabilities, its resolving power, limited by the diameter
of the telescope, is smaller than the torus size of NGC~1068 at any
given wavelength in the IR. However, \JWST\ can be used to isolate the
torus emission from potential contributions of extended diffuse dust
emission from the narrow line region (NLR), any polar/outflow
components, and star-forming regions surrounding the AGN for a large
sample of bolometric luminosities and AGN types. We also note that
10-m class telescopes, e.g., \Keck, are not able to resolve the torus
in NGC~1068 within the 1--12~\mic\ wavelength range, in line with
observations to date \citep[e.g.,][]{RamosAlmeida+2011,
  Alonso-Herrero+2011, Asmus+2015, Lopez-Rodriguez+2016}. We
henceforth focus on the observations resolved by 30~m class
telescopes.

\begin{figure*}[t]
\includegraphics[width=\textwidth,height=0.95\textheight,keepaspectratio]{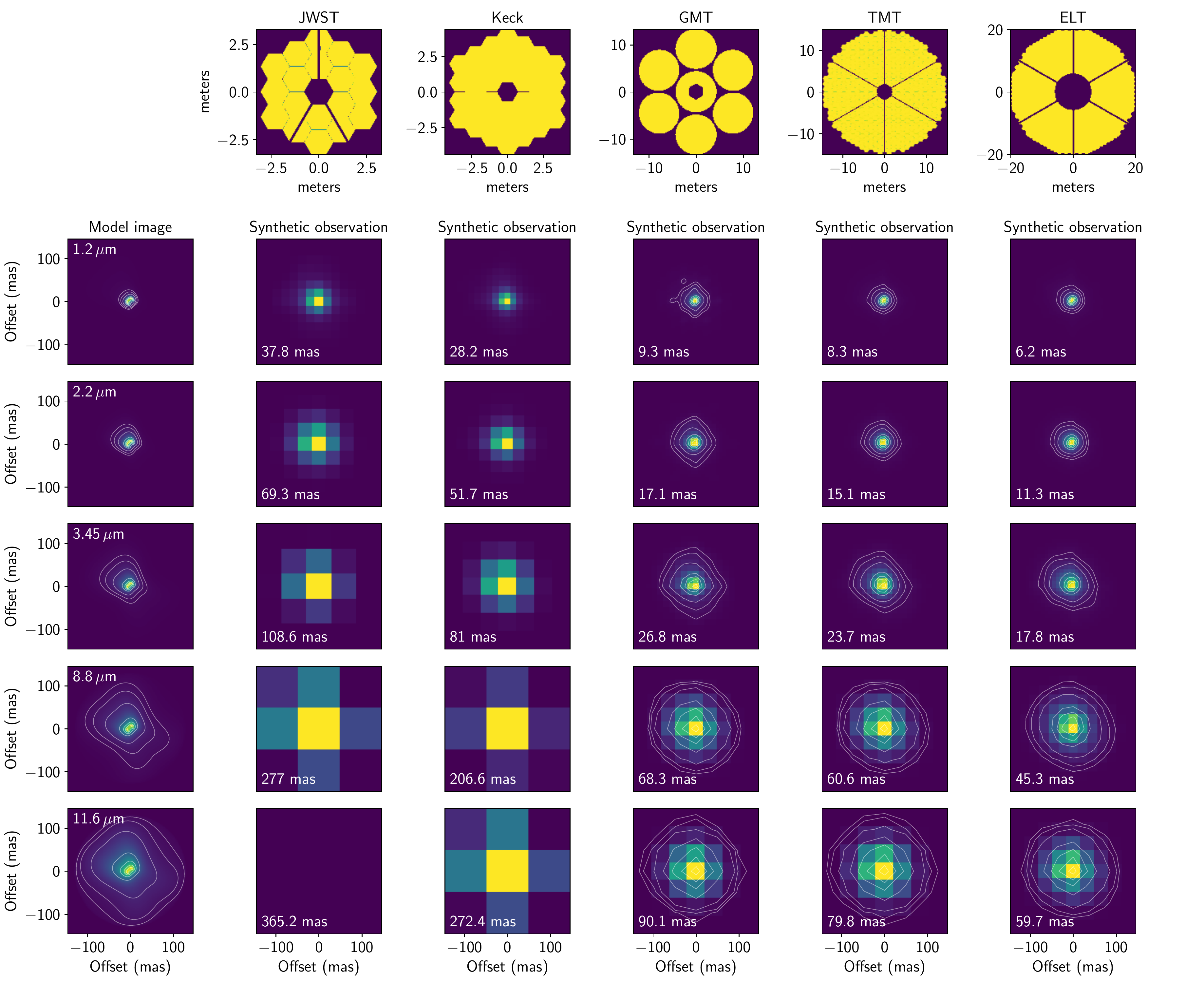}
\caption{Deconvolved observations using pixelated images from Section
  \ref{app:deconv}. First row and column as in
  Figure~\ref{fig:figI}. The other panels show the deconvolved images
  as a function of wavelength and telescope. All panels were pixelated
  to the Nyquist sampling, and print in their lower-left corners the
  angular resolution at the given wavelength estimated as
  $\lambda/D$. Contours are shown at levels of 0.03, 0.05, 0.1, 0.3,
  0.5, 0.7, 0.9 times the peak flux of the image. Contours for \JWST\
  and \Keck\ are not shown because images are consistent with point
  sources at all wavelengths. The 11.6 $\mu$m image for the \JWST\ is
  not visible because the pixel scale is similar to the FOV.}
  \label{fig:figV}
\end{figure*}

\begin{figure*}
\includegraphics[width=\textwidth,height=0.95\textheight,keepaspectratio]{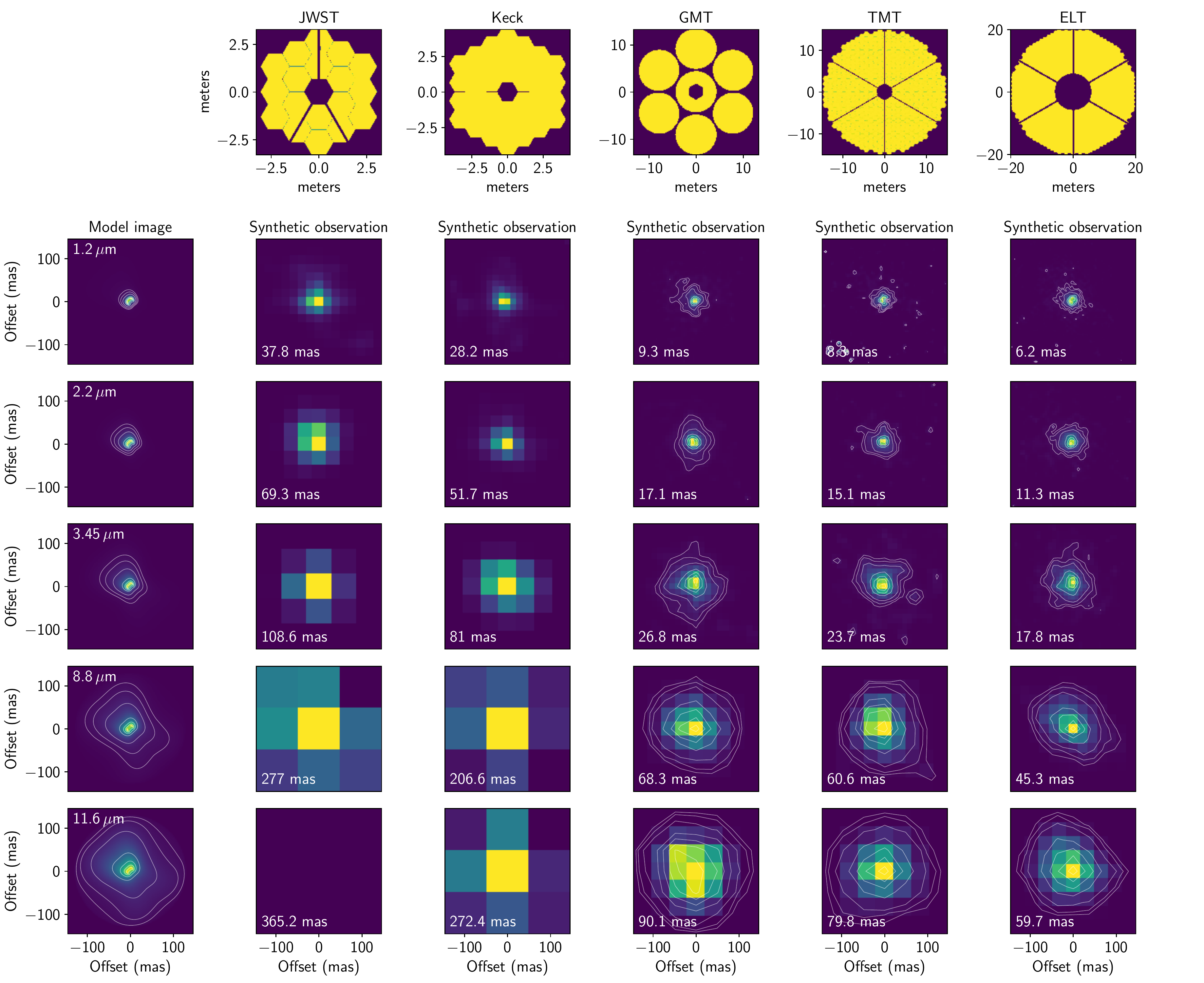}
\caption{Deconvolved observations using noisy images from Section
  \ref{app:noise}. Same as Figure \ref{fig:figV} but pixelated images
  with a SNR = 10 at the peak pixel have been deconvolved. Contours
  are shown at levels of 0.03, 0.05, 0.1, 0.3, 0.5, 0.7, and 0.9 times
  the peak of the image. Contours for \JWST\ and \Keck\ are not shown
  because images are consistent with point sources at all
  wavelengths. The 11.6 $\mu$m image for the \JWST\ is not not visible
  because the pixel scale is similar to the FOV.}
  \label{fig:figVI}
\end{figure*}

In the 1--2.5~\mic\ wavelength range the torus emission arises mostly
from the directly irradiated clouds that make up the inner walls of
the torus. In all cases, the region of emission is $<1$ pc (14 mas for
NGC~1068) which is resolved by all 30~m class telescopes. Due to the
shape of the PSF the extended torus emission is more clearly visible
with \GMT\ than with \TMT\ or \ELT. However, when deconvolution
techniques are applied, \TMT\ and \ELT\ produce resolved images that
are more comparable to the 2D original model images (see
Figures~\ref{fig:figV} and \ref{fig:figVI})

In the 3--5~\mic\ wavelength range the emission morphology becomes
more extended, and importantly, the elongation is in the polar
direction. If the torus axis is tilted with respect to the observer,
the polar component is mainly seen on one side only due to
self-obscuration by material in the equatorial region. The emission
mainly originates from the inner layers of the extended dusty torus,
and has sizes of \about 10 pc (142 mas), which are easily resolved by
all 30~m class telescopes. For all these large telescopes the core and
the extended emission are resolved. Deconvolution techniques can be
applied effectively, and are crucial in providing final images
comparable with the 2D models. These results demonstrate the
criticality of obtaining not only an excellent but also stable PSF.

In the 8--12~\mic\ wavelength range the torus emission shows an
extended component in the polar direction, with a peak of emission in
the northern inner wall of the torus. A small amount of emission is
found along the equatorial plane of the torus due to the optically
thick dust along such lines of sight. The extended emission is
marginally resolved by all the 30~m class telescopes.

Apart from telescope size, instrument configuration and sensitivity,
the physical/geometrical parameters of the AGN torus determine the
resolvability of its emission. See \paperone\ on how these parameters
affect resolvability and the existence of polar elongation of the
emission across wavelengths.


\subsection{Spatially Resolved Spectral Features}
\label{sec:ifu}
\noindent
Having the resolved images available at any wavelength, we can compute
spectral quantities per-pixel, akin to observations with an Integral
Field Unit (IFU). We demonstrate this with the 10~\mic\ silicate
feature strength \Sten, defined as
\begin{equation}
  \label{eq:s10}
  \Sten\ = \ln{\frac{F(\lambda_{10})}{F_{\rm cont}(\lambda_{10})}},
\end{equation}
i.e., the logarithmic ratio of the flux at the extremum of the 10~\mic\
feature and the continuum flux at that wavelength \citep[interpolated
as a cubic spline across the wings of the silicate
feature;][]{Spoon+2007}.

The silicates that make up the majority of ISM dust have stretching
and bending modes which produce emission at 10 and 18~\mic. In the
context of the AGN torus, the strength of the 10~\mic\ feature is a
powerful diagnostic because it is sensitive to several physical and
geometrical conditions in the torus. One of these quantities is the
inclination, which determines whether we will have a free LOS toward
the dust grains that create silicate emission. Another quantity is the
optical depth along the LOS, which, if large enough, can lead to the
silicate being seen in absorption. It is possible or even likely that
there is significant LOS silicate extinction by foreground matter,
unrelated to the torus, in many sources
\citep[e.g.,][]{RamosAlmeida+2011, Alonso-Herrero+2011, Goulding+2012,
  Gonzalez-Martin+2013, Prieto+2014, Lopez-Rodriguez+2018}.

In the context of the clumpy torus, \Sten\ is also sensitive to the
number of discrete clouds, and to their distribution in radial and
angular directions, as clumpiness fundamentally permits edge-on
configurations that afford a clear LOS toward silicate-emitting hot
dust at the far inner side of the torus. \Sten\ has been measured in
moderate absorption in type~2 Seyfert galaxies \citep{Roche+2006,
  Hao+2007, Wu+2009, Hoenig+2010, Alonso-Herrero+2016}. Deep
absorption features are a hallmark of smooth dust distributions, and
are observed in ultra-luminous infrared galaxies (ULIRGs), which are
shrouded in copious amounts of dust on large scales, but not in
Seyferts \citep{Levenson+2007}. Silicates have been observed in
emission in type~1 Seyferts \citep{Siebenmorgen+2005, Hao+2005,
  Sturm+2005, Hoenig+2010, Alonso-Herrero+2016}, even in galaxies that
were classified as type~2 for their lack of broad emission lines
\citep{Teplitz+2006, Mason+2009, Nikutta+2009}.

\begin{figure*}
  \center \includegraphics[width=\hsize]{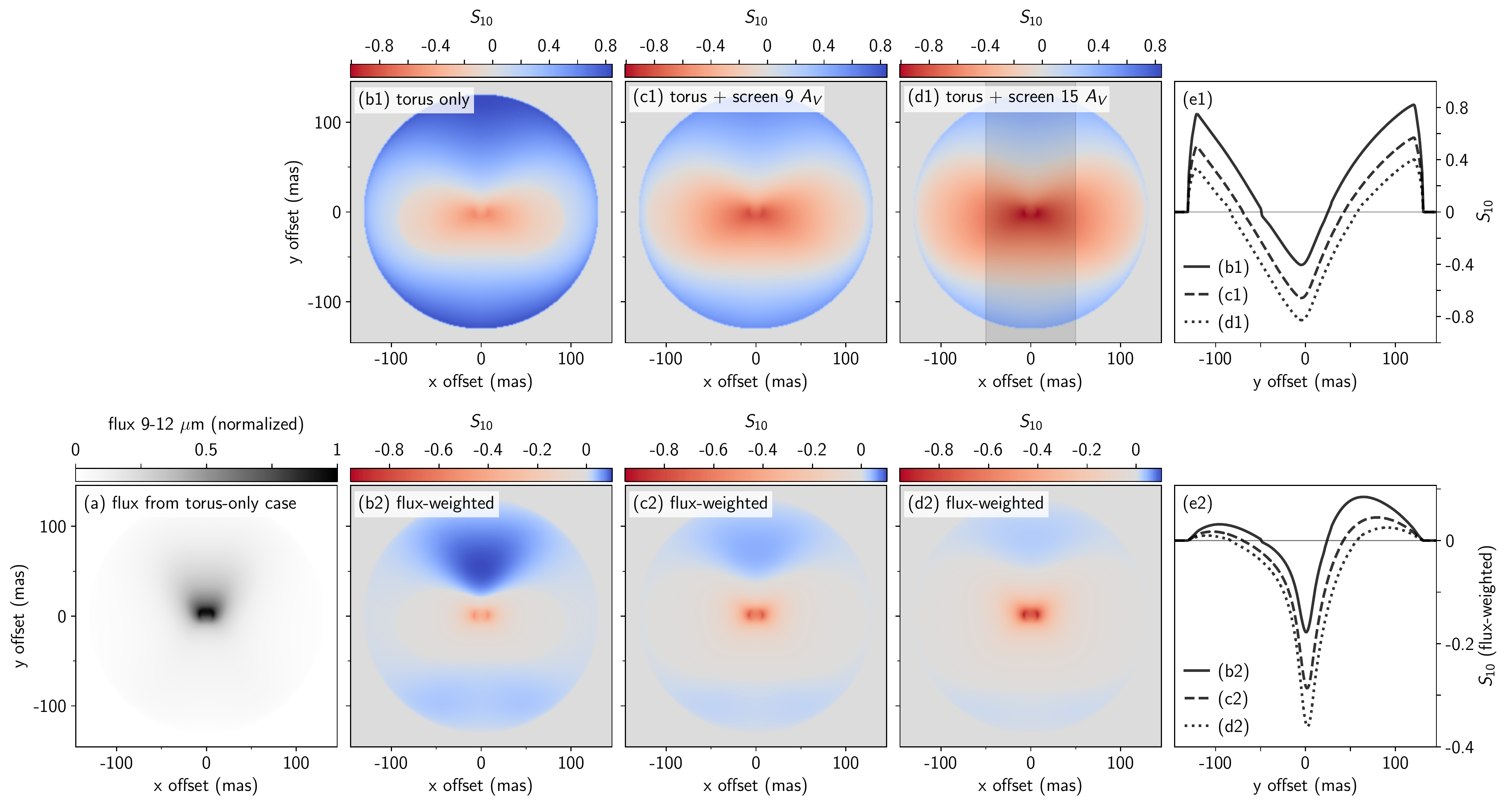}
  \caption{Spatially resolved 10~\!\mic\ silicate feature strength
    \Sten\ obtained through simulated IFU-like observations of our
    NGC~1068 model, with parameters \sig\ = 45\degr, \iv\ = 75\degr,
    \Y\ = 18, \No\ = 4, \q\ = 0.08, \tv\ = 70. Full-resolution images
    are used, and we do not consider here degradation effects such as
    PSFs, noise, atmospheric conditions, etc. We assume luminosity
    $\Lb = 1.6\times10^{45}~\rm{erg s^{-1}}$ and distance
    $D = 14.4$~Mpc. For clarity the images are not rotated to the \pa\
    of the source, i.e., the polar direction is up. \emph{(a)} Flux
    map of the torus model (no dust screen), integrated between 9 and
    12~\mic\ and normalized to its peak. \emph{(b1)} Resolved \Sten\
    map for the torus model. The torus near side produces silicate
    absorption (red colors). Polar regions generate silicate emission
    (blue colors). \emph{(c1)} \Sten\ map for a torus model behind a
    (wavelength-dependent) cold screen of ISM dust with $A_V$ = 9 mag,
    similar to results from \loro. \emph{(d1)} With a screen of $A_V$
    = 15 mag. Panels (b1)--(d1) use the same color scale. The dust
    screen lowers silicate emission, deepens silicate absorption, and
    enhances the size of the area seen in absorption. \emph{(e1)}
    Vertical mean profiles of \Sten\ over a central (vertical) band of
    $\pm 50$~mas width (indicated as a gray vertical stripe in panel
    (d1)), measured in panels b1 (solid line), c1 (dashed), and d1
    (dotted). The $x$-axis corresponds to the $y$-axes in the \Sten\
    maps. Positive \Sten\ values indicate silicates in emission,
    negative in absorption. \emph{(b2)--(d2)} The panels (b1)--(d1)
    flux-weighted with their respective 9--12~\mic\ integrated flux
    images (similar to the one shown in (a)). All three panels are
    scaled to the same color scale. \emph{(e2)} Vertical \Sten\
    profiles in (b2), (c2), (d2), measured in the same way as in
    (e1).}
  \label{fig:s10}
\end{figure*}

For the NGC~1068 model we adopted here (with parameters \sig\ =
45\degr, \iv\ = 75\degr, \Y\ = 18, \No\ = 4, \q\ = 0.08, \tv\ = 70)
Figure \ref{fig:s10} shows spatially resolved maps of \Sten, i.e., the
silicate feature strength is computed per-pixel, using model images
generated between 9 and 12~\mic. Panel (b1) shows the \Sten\ map for a
torus-only model (no intervening dust screen). In panels (c1) and (d1)
the same torus model is viewed through a screen of ISM dust (same
composition as our dusty torus clouds). We note that inhomogeneous
dust screens could alter the results we derive below due to light
leakage \citep{Krugel_2009}, but we do not consider them
here. Instead, we consider homogeneous screens with $A_V = 9$~mag of
extinction in panel (c1) \citep[similar to results
in][]{Lopez-Rodriguez+2018}, and $A_V = 15$~mag in panel (d1). The
mean profiles of \Sten\ in panel (e1), computed along the
\emph{y}-direction over a $\pm 50$~mas wide band (indicated only in
panel (d1) as a gray stripe) show how a foreground screen effectively
reduces the strength of silicate emission, deepens any silicate
absorption features, and increases the area that is observed in
absorption.

How much of the spatial \Sten\ map could be actually detected depends
on the level of flux captured by each spaxel. Figure \ref{fig:s10}
therefore also shows in panels (b2), (c2), and (d2) flux-weighted
versions of the \Sten\ maps. For the no-screen model only the central
1-2 parsec generate moderate silicate absorption (down to
$\Sten \approx -0.5$). Polar regions above and below the torus
equatorial plane produce mild silicate emission
($\Sten \lesssim 0.1$). A foreground dust screen deepens the central
silicate absorption, and dampens further the polar silicate
emission. Our modeling shows that by $A_V = 33$~mag no single spaxel
exhibits silicate emission anymore.

The behavior described above is in line with observations. For
example, \citet{Mason+2006} measure a nuclear silicate feature with
$\Sten = -0.4$. Further away from the nucleus the feature becomes more
shallow (see their figure 6). One qualification is that our model FOV
spans just about 0.4 arcseconds, i.e., comprises only the very center
of the measurements in \citeauthor{Mason+2006}. Dilution of the
central absorption feature with larger beams could lead to the
observed reduction of the weak polar emission areas. A second
difference is that in \citet{Mason+2006} the feature strength observed
along profiles just 0.4~arscec away from the nucleus is
\hbox{$\Sten \lesssim -0.2$}, i.e., the feature is still in
absorption, but weaker. This can be achieved if in addition to the
nuclear absorption caused by the torus dust there is an extrinsic
absorption screen along the line of sight, generating an additional
$\Delta \Sten \approx -0.2$. Such a screen has been invoked to explain
observed data \citep[e.g.,][]{Raban+2009, Alonso-Herrero+2011,
  Lopez-Gonzaga+2014}, and can be physically associated with dust in
the inner bar of NGC~1068.

\section{Reconstructed Interferometric Observations: Direct Image Fitting}
\label{sec:direct-image-fitting}
\noindent
\gravity\ have recently published a reconstructed K-band
image\footnote{The reconstructed K-band image used in our fitting
  routine can be found at
  \url{http://cdsarc.u-strasbg.fr/ftp/J/A+A/634/A1/fits/}} of
NGC~1068, obtained by combining the light of four telescopes with
\VLTI. In contrast with previous efforts this image was reconstructed
using phase closure information, i.e., it does not rely on nonphysical
models of the brightness distribution.

\gravity\ offered several models to explain the observed data. Their
favored model is a simple geometric ring, where the 2.2~\mic\ emission
is optically thin, and emerges even from the near side of the
ring. The authors derive a sublimation radius $R_d = 0.24$~pc,
inclined at $\iv = 70\pm5$\degr, and at a position angle of the disk
equatorial plane $\pa = -50\pm4$\degr\ (i.e., 50\degr\ west of north),
or equivalently, $\pa = 130$\degr\ E of N of the disk plane, and
consequently, $\pa = 40$\degr\ E of N for the disk rotation axis. The
scale height was constrained by the maximum beam size to $h/r<0.14$
(which would correspond to $\sigma < 8$\degr\ assuming a sharp-edge
flared disk). The bolometric luminosity is rather uncertain in the
literature, and the authors give a range of
$0.4-4.7 \times 10^{45}\, {\rm erg\,s^{-1}}$. \gravity\ also discuss
a CAT3D \citep{Hoenig_Kishimoto_2010} model (their ``Model 4''), where
the observed emission predominantly arises from the far inner side of
a disk/torus, and is therefore optically thick. This model is
ultimately disfavored in \gravity\ due to the harder-to-explain patchy
nature of the observed emission.

We fit the GRAVITY image directly with monochromatic 2.2~\mic\ \C\
model images by varying the relevant model parameters. While fitting,
we convolve the models with the effective beam (given by \gravity\ as
an $x \times y = 3.1 \times 1.1$~mas FWHM 2D Gaussian, rotated
counter-clockwise by $\pa = 48.6$\degr\ (i.e., this is the position
angle of the semi\emph{minor} axis of the Gaussian). We also account
for correct pixelization, and take advantage of the freedom to
self-similarly stretch the model image maps (``zoom factor''). In
nature the ``zooming'' can be achieved by either changing the distance
$D$ to the source -- then, angular size
\hbox{$\theta \propto {\rm D}^{-1}$} -- or by changing the AGN
luminosity; in this case $\theta \propto {\rm L}^{-1/2}$, from the
equation for the dust sublimation radius
\begin{equation}
  \label{eq:Rd}
  \Rd\ = 0.4 \left(\dfrac{\Lb}{10^{45}\, {\rm erg\, s^{-1}}} \right)^{\!1/2} \left( \dfrac{T_{\rm sub}}{1500\, {\rm K}} \right)^{\!-2.6}\, {\rm (pc)}
\end{equation}
(see also \paperone). With the fitted zoom factor we can directly
determine the luminosity \Lb\ if the source distance D is fixed, and
vice versa.

\noindent
We perform two separate fits:
\begin{enumerate}
\item Model (I): fixed-origin
\item Model (II): shifting-origin
\end{enumerate}
In model (I) the origin of the emission is fixed at the central pixel
(0,0). In model (II) the emission pattern is allowed to shift within
the image plane. This has strong implications for the resulting
geometry. Performing purely morphological fitting, with both data and
model images normalized to unit sum, we minimize a loss function
between the data and model image (RMSE, see
Appendix~\ref{sec:appendix-directimagefitting}) using a robust genetic
algorithm (differential evolution; \citealt{diff_evo}).

\begin{figure*}
  \includegraphics[width=1.\hsize]{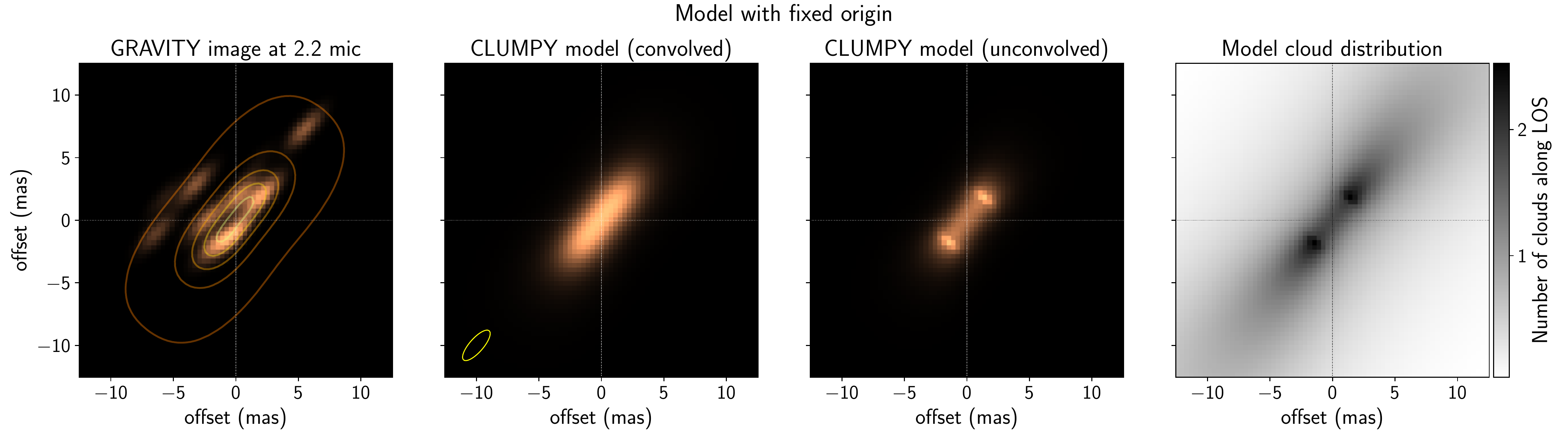}\\[10pt]
  \includegraphics[width=1.\hsize]{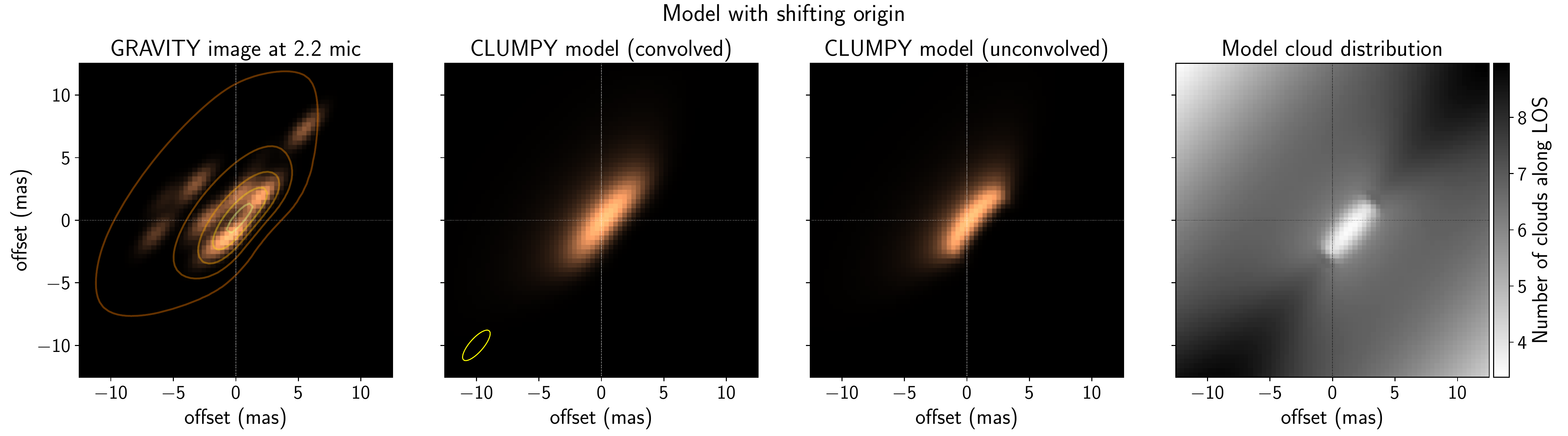}
  \caption{Fitting of the 2.2~\mic\ GRAVITY image of NGC~1068 with
    convolved \C\ images. The top row shows the best-fit fixed-origin
    model (I), and the bottom row the shifting-origin model (II). All
    emission images are normalized to unit sum. In all panels the
    dotted horizontal and vertical lines indicate the aperture origin
    at (0,0) offsets. The left panels show the central
    $25\times25$~mas image obtained by GRAVITY. The second-from left
    panels show the best-fit \C\ model images, convolved with the
    GRAVITY effective beam ellipse of FWHM = $3.1\times1.1$~mas (see
    lower-left corner of the panel). The isophotes of these model
    images are shown as contours in the left panels, at 1, 10, 30, 60,
    90 percent of the peak. The third-from-left panels show the
    underlying \C\ model images but without convolution. Finally, the
    right-most panels (in gray-scale) show the underlying dust
    distributions, each normalized to its own range of values. All
    best-fit model parameters are given in
    Table~\ref{table:gravity_fitting}.}
  \label{fig:gravity_bestfit}
\end{figure*}

The best-fit (convolved) model images at 2.2~\mic\ are shown in
Figure~\ref{fig:gravity_bestfit} in the second-from-left panels, for
the \emph{fixed-orgin} model at the top, and the shifting-origin model
at the bottom. The leftmost panels show the same GRAVITY image, with
the isophotes of the convolved models overplotted. The underlying
unconvolved model images are shown in the third-from-left
panels. Finally, the rightmost panels show the dust cloud
distributions that generate the respective emission model images,
scaled to their min/max ranges of clouds per LOS. All model images
have of course the same pixel scale as the GRAVITY reconstructed
image.

We give all best-fit parameter values in
Table~\ref{table:gravity_fitting}. Because the K-band morphology does
not depend on the torus radial extension \Y, we can fix it at an
arbitrary value and are setting it to $\Y = 18$ from SED fitting by
\loro.

\subsection{Results for the Fixed-origin Model (I)}

For the \emph{fixed-origin} model (I) all but two fitted parameters
converge robustly. The results suggest a geometrically thin structure
($\sig = 24.6$\degr\ for our Gaussian soft-edge torus), viewed at
$\iv = 85.2$\degr\ inclination, and with a fitted $\pa = 51.9$\degr\
of the torus axis (E of N, CCW), or $\pa = 141.9$\degr\ of the torus
equatorial plane. The steepness of the radial cloud number
distribution is $\propto r^{-1.35}$.
\begin{deluxetable}{lcrr}
  \tablecaption{Best-fit parameters (top) and derived quantities
    (bottom) from direct fitting of the \gravity\ NGC~1068 K-band
    image, for the fixed-origin and shifting-origin
    models.\label{table:gravity_fitting}} \tablewidth{0pt} \tablehead{
    Parameter & Notes & Fixed-origin & Shifting-origin} \startdata
  \sig\ (deg)                        &       & 24.6               & 15.1$^{a}$   \\
  \iv\ (deg)                         &       & 85.2               & 69.3         \\
  \Y                                 & (1)   & fixed at 18        & fixed at 18  \\
  \No                                &       & 1$^{a}$            & 10.9$^{b}$   \\
  \q                                 &       & 1.35               & 0.47         \\
  \tv                                &       & 10$^{a}$           & 143.5$^{b}$  \\
  \pa\ (deg)                         & (2)   & 51.9 / 141.9       & 50.3 / 140.3 \\
  x-shift (mas)                      & (3)   & n/a                & 1.37$^{c}$   \\
  y-shift (mas)                      & (3)   & n/a                & -0.74$^{d}$  \\
  zoom                               & (4)   & 1.037              & 1.344        \\
  \hline
  \Rd\ (mas / pc)                    & (5)   & 2.26 / 0.16        & 2.93 / 0.20  \\
  $\Lb\ (10^{44}\rm \,erg\, s^{-1})$ &       & 1.56               & 2.62         \\
  \enddata \tablecomments{For all conversions a distance D=14.4 Mpc
    was adopted. Notes: (1) \Y\ is almost irrelevant for the
    morphology of the 2.2~\mic\ image. (2) \pa\ of torus axis / plane,
    E of N. (3) E-W (x) and N-S (y) shift of the AGN locus w.r.t. the
    origin. Modeled in (fractional) pixel space, and converted here to
    milliarcseconds. (4) Self-similar stretch factor applied to raw
    \C\ image. (5) Angular and physical size of a dust sublimation
    radius. $^{a}$Upper bound. $^{b}$Not well-constrained. Using the
    mean of the last third of all minimization scheme iterations; see
    Appendix~\ref{sec:appendix-directimagefitting} for
    details. $^{c}$Shifted to W. $^{d}$Shifted to S.}
\end{deluxetable}
The mean number of clouds along radial equatorial rays, \No, and the
optical depth per cloud at visual, \tv, do drift toward the smallest
values present in our model hypercube, but do not appear to stabilize
before. While we can thus not constrain them further, they do occupy
the optically thin regime preferred by the \gravity\ ring model, with
the 2.2~\mic\ emission emanating also from our \emph{near} side of the
dusty disk / torus. In fact, in our best-fit model with
$\sig = 24.6$\degr, and taking $\No = 1$, there are on average only
0.96 clouds along our LOS to the central source (at
$\iv = 85.2$\degr), according to Equation 3 from \paperone. If their
optical depth at visual (0.55~\mic) is 10, then the overall LOS
optical depth in the K band is $\tau_{2.2} = 0.89$ when assuming our
standard silicate-graphite ISM dust, where
$\tau_{2.2}/\tau_{0.55} = 0.093$.

This optically thin configuration results in a K-band morphology which
very closely resembles the underlying dust distribution, apparent in
the two right panels in the top row of Figure \ref{fig:gravity_bestfit},
including the brightened edges of the central dust-free cavity. It
seems challenging to reconcile such an optically thin disk with the
amount of emission observed, and with the existence of co-planar and
co-located maser spots \citep[e.g.,][]{Gallimore+2004} which require
high densities.

The fitted zoom factor 1.037 of the images, needed in comparison to
their original resolution in the model hypercube, is small. With the
native \C\ image resolution of $6\, {\rm pix} / \Rd$ and a pixel scale
of the GRAVITY image of $0.3634\, {\rm mas / pix}$, a zoom factor of
1.037 means that the dust sublimation radius \Rd\ subtends 2.26 mas on
the sky. If we adopt the canonical distance to NGC~1068,
${\rm D} = 14.4$~Mpc, then the physical size of \Rd\ is just
0.158~pc. This is in line with some previously derived values
\citep[e.g.,][]{Burtscher+2013}, but on the lower end of measurements
reported in the literature. Equation~\eqref{eq:Rd} then yields a
sublimation-setting luminosity of
${\rm L} = 1.56\times10^{44}\, {\rm erg\,s^{-1}}$. This value is even
smaller than the lower limit from modeling by \gravity. This
apparently small luminosity may have several causes, as follows.

(i) During fitting we convolve a \C\ image with the effective beam of
the telescope/instrument configuration at the time of observation,
which \gravity\ estimate as a $3.1 \times 1.1$~mas FWHM elliptical
beam, with the short axis inclined 48.6\degr\ E from N. This
inclination is unfortunately almost identical to the \pa\ of the
2.2~\mic\ emitting dust structure; the convolution of the model image
with the beam smears out the emission \emph{along its long axis},
effectively \emph{increasing} its apparent angular size. All the while
the underlying model is smaller. In our tests this effect can increase
the apparent angular size by up to 15--20 percent.

(ii) The dust sublimation temperature also enters
Equation~\eqref{eq:Rd}, and it is often assumed at 1500~K. Increasing
T by just a few hundred Kelvin, e.g., through a slight over-abundance
of graphites, and/or by having a slightly larger average grain size,
can easily produce smaller \Rd\ at ``standard'' luminosities. For
instance, ${\rm L} = 0.4\times10^{45}\, {\rm erg\,s^{-1}}$ and
${\rm T} = 1800$~K (the sublimation temperature of pure graphite
grains, which are likely to be the ones to survive closest to the AGN)
produce $\Rd = 0.16$~pc, in line with our modeling. See, e.g.,
\citet{Mor+2009, Heymann_Siebenmorgen_2012, Xie+2015} for more
discussion on the influence of dust chemistry on the dust sublimation
temperature and sublimation radius in AGNs.

(iii) In our models the central illuminating source is isotropic. If
it were instead emitting like an accretion disk with a $\cos\theta$
intensity fall-off with polar angle $\theta$ \citep[see,
e.g.,][]{Netzer_1987}, then the dust sublimation radius is a function
of $\theta$. \citet{Kawaguchi_Mori_2010} have shown that in this case
the average \Rd\ is only about a third of the isotropic case, and
close to edge-on views it can be as small as 10 percent of the
isotropic radius. The dusty disk in NGC~1068 is seen at an angle
closer to edge-on, and therefore Equation~\eqref{eq:Rd} may not be
applicable as-is, but requires an angle-dependent luminosity
correction.

\subsection{Results for the Shifting-origin Model (II)}

For model (II) all sampled parameter ranges were kept the same, and we
allowed the 2.2~\mic\ emission pattern to move up to $\pm 10$ pixels
in the $x$ and $y$ directions within the 69-pixel wide FOV used for
fitting. A majority of the parameters have again converged
robustly. \No\ and \tv\ do not converge to a fixed value, but for
later iterations of the loss function minimization their long-term
mean is stable. For these two parameters we therefore adopt the mean
of the final third of all iterations (over 12,000 iterations),
i.e., \No\ = 10.9 clouds and \tv\ = 143.5.

The freedom to change a bit the locus of the AGN allowed for optically
thick models, where both the number of clouds along the LOS, and the
optical depth of individual clouds are higher. Such models, when
viewed at an angle somewhat above the equatorial plane (\hbox{\iv\ =
  69.3\degr} in our case), produce thermal emission morphologies that
are vertically offset from the mid-plane. For our best-fit model the
shifts amount to 1.37 and 0.74 milliarcseconds to the west and south,
respectively.

The resulting morphology, after convolution (second-from-left panel in
the lower row in Figure \ref{fig:gravity_bestfit}), is strikingly
similar to that of model (I); the RMSE is \about 6 percent lower than
in model (I). However, the rightmost panel in
Figure \ref{fig:gravity_bestfit} reveals just how different the
underlying dust distribution is in this case, and by how much it had
to be offset to cause this effect. In the unconvolved image
(third-from-left panel) it is evident how the emission emanates just
above the inner, thin-disk like part of the dust cloud distribution.

The radial cloud distribution is much flatter than in model (I), with
\hbox{\q\ = 0.47}. This is favorable for producing emission patterns
preferentially elongated in polar directions, but in this case,
together with the small value of $\sig \approx 15$\degr, is to first
order not enough to generate the MIR elongation of 2.3:1 measured for
NGC~1068 \citep{Lopez-Gonzaga+2016b}; in fact, the corresponding
10~\mic\ image of this model only produces $e \approx 0.4-0.5$ at the
0.1-contour level (of the peak). We discuss the issue of polar
elongation in detail in Section 3 of \paperone.

The zoom factor found with this shifting-origin model (II) is 1.344,
and translates to $\Rd\ = 2.93$~mas $\equiv 0.2$~pc, and
$\Lb\ = \rm 2.62 \times 10^{44}\, erg\,s^{-1}$ (all for distance D =
14.4~Mpc). The same caveats (i)--(iii) for small luminosities apply as
for model (I), but to a significantly lesser degree.

Overall the GRAVITY image is striking both in its unprecedented
spatial resolution, but also in the fact that the circumnuclear NIR
emission appears to be concentrated in only a few point-like peaks of
sufficient intensity to be detected. This calls for a cautious
interpretation. We emphasize that the instrumental beam smears out the
observations and causes them to appear very similar for different
underlying geometries. These could be either an elongated structure
suggestive of a highly inclined emission ring, or a geometrically thin
but optically thick flared disk where the emission arises from a
narrow strip of hot cloud surface layers on the far inner side of the
torus funnel. As noted in \gravity, extremely precise NIR astrometry
of future instruments could help distinguish between the two
scenarios, especially with quantitative comparison of model images
such as \C.


\section{Summary}
\label{sec:summary}

\noindent
In this paper we computed synthetic observations of the nearby AGN
NGC~1068 using images generated by the \C\ torus models
\citep{Nenkova+2008a, Nenkova+2008b}. We have obtained from the
respective observatories and consortia the pupil images of the JWST
and Keck facilities as representatives of the best current (or
nearly-current) space-based and ground-based telescopes, and of GMT,
TMT, and ELT as the upcoming generation of extremely large telescopes,
whose apertures will be in an unprecedented class of their own. From
the pupil images we compute wavelength-dependent PSFs and discuss
their general characteristics and resolving power.

We then convolve \C\ model images with these PSFs to simulate
synthetic observations of NGC~1068. We take image noise and detector
pixelization into account, and employ image deconvolution techniques
to obtain more realistic simulated observations.\footnote{The
  processing software may also be applied to analyze different
  emission models, which need not be limited to AGN tori.} Comparison
with observational results suggest that the tori in nearby AGNs, within
a range of possible luminosities and distances, will be very well
resolvable in the range $\sim$3--12~\mic\ with the new class of
extremely large telescopes, with an optimum between approximately 4
and 8 micron.

\HC\ can be used to simulate IFU-like observations. We use this
capability to generate spatially and spectrally resolved images of the
10-\mic\ silicate feature for a model of NGC~1068. We collapse the
image cubes to maps of the feature strength as function of spatial
coordinates, both without and with an intervening foreground dust
screen, which seems to be suggested by observational data. Profiles of
the silicate feature strength \Sten\ along the system polar direction
indicate that the central 1-2 parsec generate a feature in moderate
absorption, while areas above and below the torus equatorial plane
produce mild silicate emission. An intervening cold dust screen
deepens the absorption feature and dampens any emission features. We
point out that radiative models with other emission lines (e.g., CO,
\ion{H}{1}) as computed, e.g., in \citet{Wada+2016} or
\citet{Williamson+2020} can use this tool to compute synthetic IFU
observations.

We have also fit directly the model-free K-band image of NGC~1068
obtained by GRAVITY, using convolved \C\ images, and two different
regimes (fixed-origin: optically and geometrically thin; and
shifting-origin: optically thick but geometrically thin). In both
cases we were able to constrain several geometrical parameters and
find somewhat geometrically thin dusty structures seen closer to
equatorial inclinations, at position angles of the torus axis of
\about 50-52\degr\ E of N. Our fitting also yields scaling parameters
necessary to stretch the model images self-similarly, and assuming a
standard distance to NGC~1068 of 14.4~Mpc we find quite small
bolometric luminosities of
$1.6-2.6 \times 10^{44}\,{\rm erg\,s^{-1}}$. Whether this value is
robust or subject to several corrective factors, remains to be further
studied.

The question of course arises whether model images fit to a single
observed K-band morphology can constrain the model to such degree that
it reproduces observations at other wavelengths, especially in the
MIR, and whether it reproduces the broadband SED. Ideally, a fit under
morphological constraints would match photometric (SED) and
interferometric (morphology in certain bands) observations
jointly. Our preliminary Bayesian SED fits to the data assembled in
\loro\ for NGC~1068, with the addition of an extinction screen, yield
overall good fits, but the MIR elongations of the sampled models range
only between 0.96 and 1.3. While this may be sufficient to explain
elongations on scales of a few parsec, it is insufficient for the
large-scale (tens of parsec) elongations observed in NGC~1068. We
defer a full treatment of the joint fits to a future publication.

Building on this work we intend to expand our approach and simulate
observations of a larger sample of nearby and bright AGNs with
extremely large telescopes and interferometers, with the goal of
constraining the torus geometry through resolved imagery.

\C\ image hypercubes and the \HC\ software will prove useful to the
AGN field and beyond. The authors welcome inquiries and help requests,
as well as code and model contributions from the community.

\acknowledgments

{We wish to thank Leonard Burtscher, Konrad Tristram, Marko Stalevski,
Moshe Elitzur, Ric Davies and Tanio Diaz Santos for illuminating
discussions on the subjects of this paper.
We are thankful to the referee whose comments helped improve the manuscript.
R.N. acknowledges early support by FONDECYT grant No. 3140436.
E.L.-R. acknowledges support from the Japanese Society for the Promotion
of Science (JSPS) through award PE17783, the National Observatory of
Japan (NAOJ) at Mitaka and the Thirty Meter Telescope (\TMT) Office at
NAOJ-Mitaka for providing a space to work and great collaborations
during the short stay in Japan.
K.I. acknowledges support by the Program for Establishing a Consortium
for the Development of Human Resources in Science and Technology,
Japan Science and Technology Agency (JST), and partial support by the
Japan Society for the Promotion of Science (JSPS) KAKENHI (20H01939;
K.~Ichikawa).
C.P. acknowledges support from the NSF grant number 1616828.
S.F.H. acknowledges support by the EU Horizon 2020 framework programme
via the ERC Starting Grant DUST-IN-THE-WIND (ERC-2015-StG-677117).
A.A.-H. acknowledges support through grant PGC2018-094671-B-I00
(MCIU/AEI/FEDER,UE). A.A.-H. work was done under project
No. MDM-2017-0737 Unidad de Excelencia ``Mar\'{\i}a de Maeztu'' -
Centro de Astrobiolog\'{\i}a (INTA-CSIC).
R.N., E.L.-R., K.I. are very grateful to NOAO (now part of NSF's NOIRLab),
SOFIA Science Center, and to the Program for Establishing a Consortium
for the Development of Human Resources in Science and Technology,
Japan Science and Technology Agency (JST), for providing travel grants
that made three on-site project meetings possible.
We are grateful to colleagues at several telescope collaborations and
consortia who were willing and able to provide us with the latest
versions of their pupil images. These are, in order of increasing
telescope diameter: Marshall Perrin (\JWST), Andrew Skemer (\Keck),
Warren Skidmore and Christophe Dumas (\TMT), Rebecca Bernstein (\GMT),
and Suzanne Ramsey (\ELT). With their permission we are here
publishing these pupil images as FITS files (see Supplements in
\paperone, and the project repository
\url{https://github.com/rnikutta/hypercat/}).}\\

\facilities{
  \JWST, \Keck, \VLTI\ (GRAVITY)
}

\software{
  \HC\ \citep{hc-paper1}, \texttt{astropy}
  \citep{astropy:2013,astropy:2018}, \texttt{h5py} \citep{h5py},
  \texttt{matplotlib} \citep{matplotlib}, \texttt{numpy}
  \citep{numpy}, \texttt{scipy} \citep{scipy}
}

\newpage

%
\bibliography{paper}{}
\bibliographystyle{aasjournal}

%

\appendix

\section{Synthetic Observations with Single Dish Telescopes}
\label{app:synobs}
\subsection{Generate a PSF}
\label{app:PSF}
\noindent
\HC\ provides three methods to estimate the PSF for a given telescope and wavelength.\\

\noindent
{\bf Model-PSF} (\emph{Airy function + Gaussian}):
\noindent
The simplest PSF model of a circular telescope aperture is an Airy
disk, i.e., the Fourier transform of the aperture. It is sufficient for
seeing-limited observations. Focusing on 30~m class and space-based
telescopes, our modeled observations will be nearly diffraction
limited in the near- to mid-IR. A better PSF model is then a
combination of a central Airy disk, which contains most of the power
of the PSF, and a broad halo that encompasses the remaining power of
the object \citep[e.g.,][]{Hardy_Thompson_2000}. This is a
semi-empirical function determined by fitting the PSF of adaptive
optics observations at optical and NIR wavelengths. The free
parameters of the model are the primary mirror diameter $D$, the
Strehl ratio $S$, and the wavelength of the observation $\lambda$.

The full-width at half-maximum of the Airy core
${\rm FWHM_A} = \lambda/D$ is defined as the location of the first
minimum such that
\begin{equation}
  {\rm PSF_A} = I^{A}_{0} J_{1} ({\rm FWHM_A}/2)
\end{equation}
where $I^{A}_{0}$ is the peak of the core of the PSF and $J_{1}$ is
the normalized first order Bessel function used by the 2D Airy
function from \textsc{astropy}.

The normalized intensity of the peak `p', for a Strehl ratio S$_{p}$,
is defined as $I^{A}_{0} = e^{-\sigma_{p}^{2}}$
\citep[see][]{Hardy_Thompson_2000}, where $\sigma_{p}$ is the
mean-square wavefront error. For typical diffraction-limited
observations $\sigma_{p_{0}} = 0.0745$ at a Strehl of
$S_{p_{0}} = 0.8$. Using these values as normalization factors of the
mean-square wavefront error,
$\sigma_{p_{0}}/S_{p_{0}} = 0.0745/0.8 = 0.093125$, and we can then
define the peak of the core as a function of the Strehl ratio,
$S_{p}$, as $I^{A}_{0} = e^{-(0.093125/S_{p})^{2}}$. Note that the
peak of PSF$_{A}$ is always $<1$ because the missing flux is
distributed within the halo (Gaussian profile), which is taken into
account in Equation \ref{eq:Ig}.

\begin{figure}
  \begin{centering}
  \includegraphics[width=\hsize]{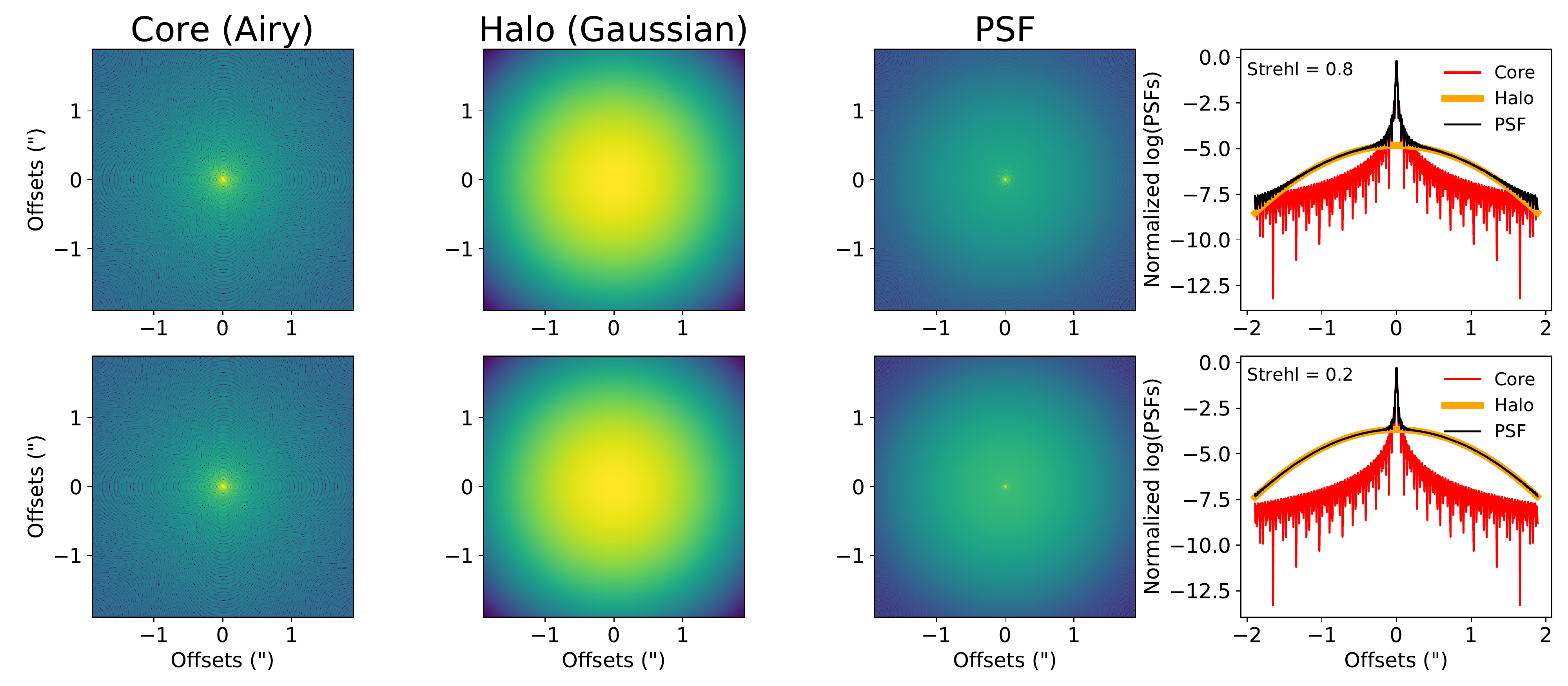}
  \caption{Model-PSF for a 30~m telescope at 2.2 \mic~with a
    $\rm FWHM_{A} = 15$ mas and Strehl ratios of 0.8 (top) and 0.2
    (bottom). From left to right the panels show the step-by-step
    generation of the model-PSF, using as core an Airy profile (first
    column), a Gaussian profile for the halo (second column), and the
    combined model-PSF (third column). A radial profile (forth column)
    for each PSF is shown. Larger Strehl ratios result in more power
    in the core.}
  \label{fig:figIII}
  \end{centering}
\end{figure}

In the general 2D case the halo is given by a double Gaussian profile
\begin{equation}
  {\rm PSF_G} = I^G_0 \exp{\left[- 4 \ln 2 \left\{ \left( \frac{x-x_0}{\rm FWHM_{\,G}^{\,x}}\right)^{\!2} + \left( \frac{y-y_0}{\rm FWHM_{\,G}^{\,y}}\right)^{\!2} \right\} \right]},
\end{equation}
which in the circular case simplifies to
\begin{equation}
  {\rm PSF_G} = I^G_0 \exp{\left[-\frac{4 \ln 2}{{\rm FWHM_{\,G}^{\,2}}} \left\{ (x-x_0)^2 + (y-y_0)^2 \right\} \right]}.
\end{equation}
$I^{G}_{0}$ is the peak of the Gaussian profile
\begin{equation}
  I^{G}_{0} = \frac{1-e^{-(0.093125/S_{p})^{2}}}{1 + \left(\dfrac{D}{\rho_0}\right)^{\!2}}.
  \label{eq:Ig}
\end{equation}
The peak of the Gaussian profile accounts for the power missed by the
central PSF, where $I_{0}^{A} + I^{G}_{0} = 1$.
${\rm FWHM_G}$ is defined as
\begin{equation}
  {\rm FWHM_G} = 1.22\,\frac{\lambda}{D}\left[1 + \left(\frac{D}{\rho_{0}}\right)^{\!2} \right]^{1/2}.
\end{equation}
The coherent length of turbulence for short exposures is
\begin{equation}
  \rho = \rho_{0}\left[ 1 + 0.37 \left(\frac{\rho_{0}}{D}\right)^{\!1/3}\right],
\end{equation}
where $\rho_{0}$ is $\propto \lambda^{6/5}$, with typical values of
$0.1-0.2$ m at 0.5 \mic. We use the normalization of 0.15 m at 0.5
\mic. The final Model-PSF is computed as $\rm PSF = PSF_A + PSF_G$
with a normalized total peak flux. Figure~\ref{fig:figIII} shows an
example of a PSF modeled this way for a 30~m telescope at 2.2 \mic,
$\rm FWHM_A$ of 15 mas and Strehl ratios of 0.8 and 0.2.  As expected,
for larger Strehl ratios more power is accumulated in the core.\\

\noindent {\bf Pupil-PSF} (\emph{Fourier transform of the pupil}):
Several telescope structures such as the central obscuration of the
secondary mirror, the thick spiders holding it, edges of the aperture,
as well as the segmentation of the primary mirror for large
telescopes, create a PSF with features far from the ideal case
described above. To model the effects of these geometrical features on
the PSF we can use the pupil image of the telescope. The pupil image
has the shape of the primary mirror and the rest of telescope
structures in front of it. The pupil images of all telescopes in
Table~\ref{table:psfmodeling} are available as FITS files in
\HC.\footnote{Telescope pupils and metadata can be found at
  \url{https://github.com/rnikutta/hypercat/tree/master/data/}}. Where
they are available, \HC\ uses these pupil images by
default. \ref{fig:figII} shows the pupil images and the PSFs of the
telescopes from Table \ref{table:listinstruments}. Figures
\ref{fig:figII} through \ref{fig:figVI} show in their upper rows the
pupil images of several telescopes. The PSF of a real telescope can
then be estimated as the Fourier transform of the pupil image
\begin{equation}
  {\rm PSF}(x',y') = \int {\rm Pupil}(x,y)\,e^{-j2\pi(x'x +y'y)}\, \dif x\, \dif y
\end{equation}
where $(x,y)$ and $(x',y')$ are the coordinates of the object and PSF
plane, respectively, and the integrals are over the image domain. The
PSF plane, $(x',y')$, is then scaled to the pixel scale in the plane
of the sky at a given wavelength as $\lambda/D$, in units of
arcsec. Finally, \HC\ also generates a pixelated PSF with the same
pixel scale of the source. \\

\noindent {\bf Image-PSF} (\emph{PSF image provided externally}): 
The previous two models might lack additional components of the PSF. Some telescopes
provide forward-modeled PSF images at specific wavelengths, e.g., PSFs
for \textit{HST} computed with the \textsc{Tiny
  Tim}\footnote{\url{http://www.stsci.edu/hst/instrumentation/focus-and-pointing/focus/tiny-tim-hst-psf-modeling}}
tool \citep{Krist+2011}, or for \JWST\ with the
\textsc{WebbPSF}\footnote{\url{http://www.stsci.edu/jwst/science-planning/proposal-planning-toolbox/psf-simulation-tool}}
tool \citep{Perrin+2012}. Telescopes also take into account
observations of point sources to characterize their PSFs at several
instrument configurations. \HC\ can take such PSF images, together
with a pixel scale, and apply them to the 2D \C\ model images. A
drawback of this method is that a PSF image must be provided at the
exact wavelength of interest (the wavelength of the model
image). Interpolation of PSF images between sampled wavelengths is not
advisable as the PSF behavior can be very complex, and interpolation
artifacts may be of
larger magnitude than the differences between neighboring wavelengths.\\

This method accepts the PSF taken by a given instrumental
configuration can be used to produce synthetic observations. The PSF
should be provided in \textsc{FITS} format, where the header must have
the pixel scale in units of arcsec per pixel. To compute the
convolution with the model at full resolution, the PSF must also have
the same FOV as the model image. Table \ref{table:psfmodeling} shows
the available PSF modes for each telescope.
\begin{deluxetable}{lccc}
  \tablecaption{Available PSF Modeling Modes.\label{table:psfmodeling}}
  \tablewidth{0pt}
  \tablehead{
    \colhead{Telescope}  & \multicolumn3c{Mode} \\
    \colhead{(by size)}  & \colhead{model-PSF} & \colhead{pupil-PSF} & \colhead{image-PSF}
    }
    \startdata
    \JWST		& $\circ$  	& $\bullet$    	& $\circ$ \\
    Gemini     	& $\circ$  	& -             & $\circ$ \\
    \Keck       & $\circ$	& $\bullet$     & $\circ$ \\
    VLT         & $\circ$  	& -             & $\circ$ \\
    \GMT        & $\circ$ 	& $\bullet$     & - 	  \\
    \TMT        & $\circ$  	& $\bullet$    	& - 	  \\
    \ELT        & $\circ$  	& $\bullet$    	& $\circ$ \\
    generic     & $\circ$  	& $\circ$     	& $\circ$ \\
    \enddata
  \tablecomments{ $\bullet$ default\hspace{10pt} $\circ$ available\hspace{10pt} - not available}
\end{deluxetable}

\subsection{PSF Convolution}
\label{app:PSFConvolution}
\noindent
Let $I_\lambda^{\rm mod}(x',y')$ be the 2D \C\ model torus image at a
given wavelength $\lambda$, at the full spatial sampling provided by
\HC\ (as described in Section 2.5 of \paperone).
${\rm PSF}_\lambda(x',y')$ be the telescope PSF (Section
\ref{app:PSF}) at the same wavelength. Then the synthetically observed
image is given by the convolution of the full resolution model and the
PSF
\begin{equation}
  I_\lambda^{\rm obs}(x,y) = I_\lambda^{\rm mod}(x',y') \ast {\rm PSF}_\lambda(x',y') = \int I_\lambda^{\rm mod}(x',y')\, {\rm PSF}_\lambda(x-x',y-y')\,\dif x'\,\dif y' 
\label{eq:PSFconvolution}
\end{equation}
where the PSF has been normalized to unity
\begin{equation}
  \int\limits_{-\infty}^{\infty} \int\limits_{-\infty}^{\infty} {\rm PSF_\lambda}(x,y)\,\dif x\,\dif y = 1
\label{eq:PSFnormalized}
\end{equation}
and has the same spatial sampling as the 2D \C\ torus image,
$I_\lambda^{\rm mod}(x',y')$. The User Manual shows in detail how the
pupil images and PSFs can be incorporated into a workflow.

\subsection{Detector Pixelization}
\label{app:DecPixelization}
\noindent
While the PSF-convolved image is at the same (high) angular resolution
as the model image, the detector can only generate discrete elements
of resolution given by its pixel scale. We produce a pixelated image
given the instrumental configuration shown in Table
\ref{table:listinstruments}. Specifically, we down-sample
the PSF-convolved image to the desired detector pixel scale using a
spline interpolation of order 3. The PSF-convolved image with a
$N_{x}\times N_{y} = 241\,{\rm pix} \times 241\,{\rm
  pix}$ and pixel
scale $\Delta {\rm px_{full}}$ is then pixelated to achieve the
detector pixel scale $\Delta {\rm px_{det}}$, by using the pixel
conversion
$\Delta {\rm px_{obs}} = \Delta {\rm px_{\rm det}} / \Delta {\rm
  px_{full}}$. The final pixelated image has
$N'_{x}\times N'_{y} = N_{x} / \Delta {\rm px_{obs}} \times N_{y}
/\Delta {\rm px_{obs}}$ pixels$^{2}$. Figure~\ref{fig:stepbystep}
(middle row) shows an example of this procedure using the 2D
PSF-convolved \C\ model image of NGC~1068 obtained in
Appendix~\ref{app:PSFConvolution}, and pixelated to the Nyquist
sampling of \TMT\ at 3.45~\mic.

\subsection{Noise}
\label{app:noise}
\noindent
In a real measurement noise must be considered due to, for instance,
detector fluctuations, atmospheric emission and transmission, etc. In
\HC\ we approximate the (optional) noise contribution as
background-limited. The added noise pattern is a Gaussian with mean
$\mu=0$ and standard deviation $\sigma = f \times \rm max(image)/SNR$,
that is, the noise level is specified by the desired SNR, for a signal
given as fraction $f$ of the peak pixel value. The default fraction is
$f = 1$, i.e., the actual peak pixel value, but it can be set within
$\left]0,1\right]$. This is useful for instance if the observer is
interested in the SNR in the extended emission of the source, rather
than at its peak. Figure~\ref{fig:stepbystep} panel G shows an example
of this procedure using the convolved and pixelated synthetic
observations of NGC~1068 with \TMT\ at 3.45~\mic\ (see
Section~\ref{app:DecPixelization}), with a SNR of 20 at the peak pixel
(and noise level fraction $f = 1$).

\subsection{Deconvolution}
\label{app:deconv}
\noindent
Deconvolution techniques make it possible to remove the PSF
contribution from (semi-)resolved observations to obtain finer details
of the astrophysical object. The Richardson-Lucy algorithm
\citep{Richardson_1972, Lucy_1974} is one of the most commonly used
methods for image deconvolution in astronomy. We use it on the
synthetic observations generated by \HC, both for the pixelated
(Section~\ref{app:DecPixelization}) and the noisy (Section~\ref{app:noise})
images. The user can introduce the number of iterations required to
produce a final deconvolved image based on their own
criteria. Figures~\ref{fig:figV} and \ref{fig:figVI} show examples of
this procedure using the pixelated synthetic and noisy observations of
NGC~1068 with 10 iterations. For larger iterations the algorithm
produces artifacts at the lower surface brightness, while for small
iterations the algorithm did not have time to converge. Our criterion
was to stop the iterations once the final image did recover the
overall shape of the contour at the level of 0.01 $\times$ the peak flux of the
full resolution image shown in Figure \ref{fig:figV}.

\section{Direct Image Fitting}
\label{sec:appendix-directimagefitting}
\begin{figure}
  \center
  \includegraphics[width=\hsize]{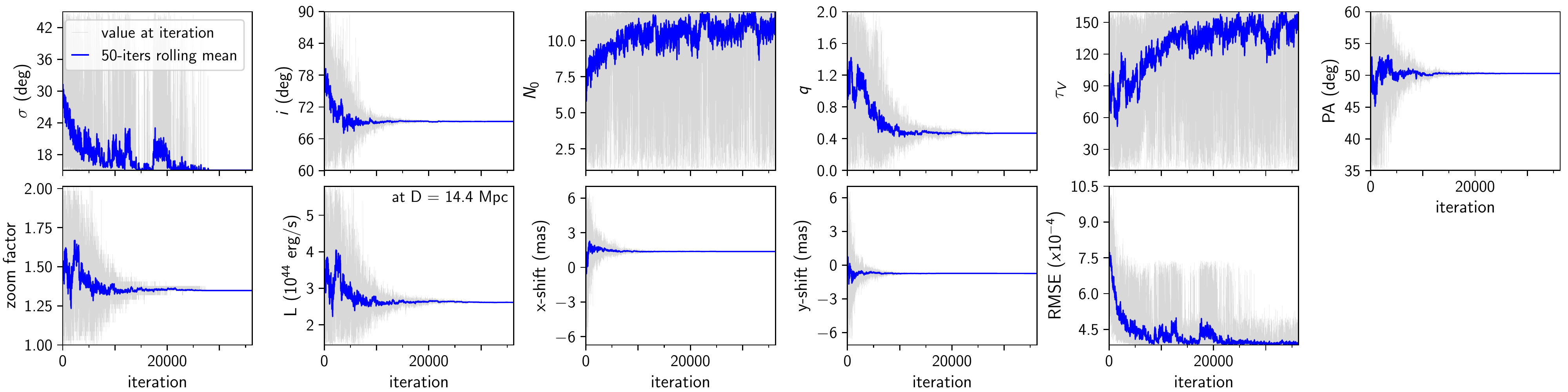} 
  \caption{Fitting of the 2.2~\mic\ GRAVITY image with convolved \C\
    images, for the case of the shifting-origin model (II); see
    Section~\ref{sec:direct-image-fitting} for details. The panels show
    the convergence behavior of the differential evolution
    minimization, with a Root Mean Square Error (RMSE, last panel)
    loss function. The \Y\ parameter was fixed at $\Y\ = 18$. $L$ is
    computed from the zoom factor after the fitting, and assuming a
    distance D = 14.4 Mpc to NGC~1068.}
  \label{fig:gravity_convergence}
\end{figure}

\noindent
The direct image fitting described in
Section~\ref{sec:direct-image-fitting} minimizes a loss function, in our
case the Root Mean Square Error
\begin{equation}
  \label{eq:rmse}
  RMSE = \sqrt{\langle(D-M)^2\rangle}\,,
\end{equation}
where $\langle\cdot\rangle$ indicates the mean over the squared
residuals between data $D$ and model $M$ (pixel values of each). We
minimize Equation~\eqref{eq:rmse} using \emph{differential evolution},
implemented in \texttt{scipy.optimize.differential\_evolution}. This
genetic algorithm \citep{diff_evo} proves very robust, is reasonably
fast even in high dimensions, and is easily parallelizable. Figure
\ref{fig:gravity_convergence} shows the convergence behavior over the
iteration number for our shifting-origin model (II).

\label{lastpage}
\end{document}